\documentclass[apj]{emulateapj}
\usepackage{mathptmx}
\usepackage{color}
\usepackage{subfigure}

\journalinfo{Draft version \today}
\slugcomment{ApJ accepted}
\shorttitle{WEAK EMISSION-LINE QUASARS AT LOW REDSHIFT}
\shortauthors{WU ET AL.}

\def\simgt{\lower 2pt \hbox{$\, \buildrel {\scriptstyle >}\over {\scriptstyle \sim}\,$}}
\def\simlt{\lower 2pt \hbox{$\, \buildrel {\scriptstyle <}\over {\scriptstyle \sim}\,$}}
\def\chandra{{\it Chandra}}

\def\rosat{{\it ROSAT}}

\def\xmm{\hbox{\it XMM-Newton}}

\def\pl{\hbox{power-law}}

\def\civ{C~{\sc iv}}
\def\mgii{Mg~{\sc ii}}
\def\lyanv{Ly$\alpha$ + N~{\sc v}}

\def\aox{$\alpha_{\rm ox}$}
\def\daox{$\Delta\alpha_{\rm ox}$}
\def\aro{$\alpha_{\rm ro}$}

\def\xray{\hbox{X-ray}}
\def\phl{PHL~1811}

\begin{document}

% -----------------------------------------------------------------------------
% Title
% -----------------------------------------------------------------------------

\title{X-ray and Multiwavelength Insights into the Nature of Weak Emission-Line Quasars \\ 
at Low Redshift}

% -----------------------------------------------------------------------------
% Authors and their affiliations
% -----------------------------------------------------------------------------

\author{Jianfeng~Wu\altaffilmark{1,2},
W.~N.~Brandt\altaffilmark{1,2},
Scott~F.~Anderson\altaffilmark{3},
Aleksandar~M.~Diamond-Stanic\altaffilmark{4,5},
Patrick~B.~Hall\altaffilmark{6},
Richard~M.~Plotkin\altaffilmark{7},
Donald~P.~Schneider\altaffilmark{1},
Ohad~Shemmer\altaffilmark{8}
%Paul~S.~Smith\altaffilmark{8}
}

\altaffiltext{1}
    {Department of Astronomy \& Astrophysics, The Pennsylvania State
     University, 525 Davey Lab, University Park, PA 16802, USA}
\altaffiltext{2}
    {Institute for Gravitation and the Cosmos, The Pennsylvania State
     University, University Park, PA 16802, USA}
\altaffiltext{3}
    {Department of Astronomy, University of Washington, Box 351580,
     Seattle, WA 98195, USA}
\altaffiltext{4}
    {Center for Astrophysics and Space Sciences, University of California,
     San Diego, La Jolla, CA, 92903, USA}
\altaffiltext{5}
    {Center for Galaxy Evolution Fellow}
\altaffiltext{6}
    {Department of Physics \& Astronomy, York University, 4700 Keele Street,
     Toronto, ON M3J 1P3, Canada}
\altaffiltext{7}
    {Astronomical Institute ``Anton Pannekoek,'' University of Amsterdam,
     Science Park 904, 1098 XH, Amsterdam, The Netherlands}
\altaffiltext{8}
    {Department of Physics, University of North Texas, Denton, TX 76203, USA}
%\altaffiltext{8}
%    {Steward Observatory, University of Arizona, 933 North Cherry Avenue, Tucson,
%     AZ 85721, USA}
\email{jfwu@astro.psu.edu}

% -----------------------------------------------------------------------------
% Abstract
% -----------------------------------------------------------------------------

\begin{abstract}

We report on the \xray\ and multiwavelength properties of 11 radio-quiet quasars with weak 
or no emission lines identified by the Sloan Digital Sky Survey (SDSS) with redshift
$z=0.4$--$2.5$. Our sample was selected from the Plotkin et~al. catalog of 
radio-quiet, weak-featured AGNs. The distribution of relative \xray\ brightness for our
\hbox{low-redshift} weak-line quasar (WLQ) candidates is significantly different from that of 
typical radio-quiet quasars, having an excess of \xray\ weak sources, but it is consistent with 
that of high-redshift WLQs. Over half of the \hbox{low-redshift} WLQ candidates are \xray\
weak by a factor of $\gtrsim 5$, compared to a typical SDSS quasar with similar
UV/optical luminosity. These \xray\ weak sources generally show similar UV emission-line
properties to those of the \xray\ weak quasar \phl\ (weak and blueshifted high-ionization lines, 
weak semi-forbidden lines, and strong UV Fe emission); they may belong to the notable class of
\phl\ analogs. The average \xray\ spectrum of these sources is somewhat harder than that 
of typical radio-quiet quasars. Several other \hbox{low-redshift} WLQ candidates have
normal ratios of \xray-to-optical/UV flux, and their average \xray\
spectral properties are also similar to those of typical radio-quiet quasars. The \xray\
weak and \xray\ normal WLQ candidates may belong to the same subset of quasars having
high-ionization ``shielding gas'' covering most of the wind-dominated broad
emission-line region, but be viewed at different inclinations. The mid-infrared-to-X-ray 
spectral energy
distributions (SEDs) of these sources are generally consistent with those of typical SDSS
quasars, showing that they are not likely to be BL~Lac objects with relativistically
boosted continua and diluted emission lines. The mid-infrared-to-UV SEDs of most radio-quiet
weak-featured AGNs without sensitive X-ray coverage (34 objects) are also consistent with 
those of typical SDSS quasars. However, one source in our \xray\ observed sample is
remarkably strong in \hbox{X-rays}, indicating that a small fraction
of \hbox{low-redshift} WLQ candidates may actually be BL~Lacs residing in the radio-faint tail of
the BL~Lac population. We also investigate universal selection criteria for WLQs 
over a wide range of redshift, finding that it is not possible to select WLQ candidates 
in a fully consistent way using different prominent emission lines (e.g., Ly$\alpha$, 
\civ, \mgii, and H$\beta$) as a function of redshift. 

\end{abstract}


% -----------------------------------------------------------------------------
% Introduction
% -----------------------------------------------------------------------------

\section{Introduction}\label{intro}

Strong and broad line emission is a common feature of quasar spectra in
the optical and UV bands. However, since multi-color quasar selection
at high redshift in the Sloan Digital Sky Survey
(SDSS; York et~al. 2000) is mostly based upon the presence
of the Ly$\alpha$ forest and Lyman break (e.g., Richards et~al.\ 2002), the SDSS can
also effectively select high-redshift quasars with weak or no emission lines.
About 90 such weak-line quasars (WLQs) at high redshift
have been found with \lyanv\ rest-frame equivalent widths of
\hbox{REW~$<15$~\AA} (e.g., Fan et~al.\ 1999, 2006; Anderson et~al.\ 2001;
Collinge et~al.\ 2005; Diamond-Stanic et~al.\ 2009, hereafter DS09). Some of
these objects show a hint of weak Ly$\alpha$ emission but no other
lines; others are completely bereft of detectable emission
lines even in high-quality spectra. High-redshift SDSS quasars show
an approximately log-normal distribution of \lyanv\ REW
with a mean of $\approx 62$~\AA\ (DS09). The WLQs constitute $\simgt 3\sigma$
negative deviations from the mean, and there is no corresponding
population with $\simgt 3\sigma$ positive deviations. The majority
of these high-redshift WLQs are radio quiet ($\alpha_{\rm ro}>-0.21$;
\aro\ is the slope of a nominal power law between 5~GHz and 2500~\AA\ in the
rest frame; see \S\ref{xray} for a full definition).

WLQs have mainly been studied at high redshifts due to the fact that 
the Ly$\alpha$ forest enters into the SDSS 
spectroscopic coverage for quasars at $z>2.2$. However, there is no apparent
reason to believe that these objects should not also exist at
lower redshifts. Indeed, a few apparent analogs of WLQs at lower
redshifts have been found serendipitously over the past $\approx
15$ years; e.g., PG~1407+265 (McDowell et~al.\ 1995;
\hbox{$z=0.94$}), 2QZJ2154--3056 (Londish et~al.\ 2004;
\hbox{$z=0.49$}), and PHL~1811 (Leighly et~al.\ 2007ab;
\hbox{$z=0.19$}). As a byproduct of a systematic survey for
optically selected BL~Lacertae objects (hereafter BL Lacs) in 
SDSS Data Release~7 (DR7; Abazajian et~al. 2009), Plotkin et~al.
(2010a) discovered about 60 additional radio-quiet WLQ candidates
at $z<2.2$ for which all emission features have \hbox{REW~$<5$~\AA}. These
objects are perhaps the first \hbox{low-redshift} SDSS
counterparts of the previously identified high-redshift SDSS
WLQs. Following the nomenclature that has been established by previous work 
on WLQs (e.g., Shemmer et~al. 2009), we define ``high redshift'' as $z>2.2$ and 
``low-redshift'' as $z\leqslant 2.2$, because WLQs are selected with different approaches
for these redshift ranges (see above). Although WLQs are rare, their exceptional characteristics
constitute a challenge to our overall understanding of quasar
geometry and physics, especially the quasar broad emission-line
region (BELR). Analogously, physical insights have been gained by
investigating other minority populations with exceptional
emission-line or absorption-line properties, such as Narrow-Line
Seyfert~1 (NLS1) galaxies and Broad Absorption Line (BAL) quasars.
Therefore, extensive studies of the multi-band properties of WLQs
should have scientific value.

There are several candidate explanations for the physical nature of WLQs.
Their UV emission lines may be weak due to an ``anemic'' BELR with a significant
deficit of line-emitting gas (e.g., Shemmer et al. 2010). It has also been speculated
that WLQs may represent an early stage of quasar evolution in which
an accretion disk has formed and emits a typical continuum, but
BELR formation is still in progress (e.g., Hryniewicz et~al. 2010;
Liu \& Zhang 2011).

The weak UV emission lines may also be a consequence of a 
spectral energy distribution (SED) which lacks
high-energy ionizing photons. This soft SED may be a result of
unusual accretion rate. For example, an extremely high accretion
rate might produce a UV-peaked SED (e.g., Leighly et~al. 2007). In
this scenario, high-ionization lines, like \civ, should be
suppressed relative to low-ionization lines like H$\beta$.
However, Shemmer et~al. (2010) estimated the normalized accretion
rates, $L/L_{\rm Edd}$, of two high-redshift WLQs via
near-infrared spectroscopy and found their accretion rates were 
within the range for typical quasars with similar
luminosities and redshifts. Alternatively, a combination of low
accretion rate and large black hole mass may lead to a relatively
cold accretion disk that emits few ionizing photons. Laor \& Davis
(2011) predicted a steeply falling SED at $\lambda\;<\;1000$~\AA\
for quasars with cold accretion disks, and such an SED was
observed in the WLQ SDSS~J0945+1009 by Hryniewicz et~al. (2010).

High-energy ionizing photons (including \hbox{X-rays}) may be
heavily absorbed before they reach the BELR. Wu et~al. (2011)
studied a population of \xray\ weak quasars with unusual UV
emission-line properties like those of \phl\ (weak and highly
blueshifted high-ionization lines, weak semi-forbidden lines, and
strong UV Fe emission). All of their radio-quiet \phl\ analogs
were found to be \xray\ weak by a factor of $\approx13$ on
average. These objects also show a harder average \xray\ spectrum than
those for typical quasars which suggests the presence of \xray\
absorption. \phl\ analogs appear observationally to be a
significant subset ($\approx 30\%$) of WLQs. The existence of a class of
quasars with high-ionization ``shielding gas'' covering most of
the BELR, but little more than the BELR, could potentially unify
the \phl\ analogs and WLQs via orientation effects (see \S4.6 of
Wu et~al. 2011). The shielding gas would absorb high-energy
ionizing photons before they reach the BELR, resulting in weak
high-ionization emission lines. When such a quasar is observed
through the BELR and the shielding gas, a \phl\ analog would be
seen; when it is observed along other directions, an \xray\ normal
WLQ would be observed.

Another possibility is that instead of being intrinsically weak,
the UV emission lines of WLQs could in principle be diluted by a
relativistically boosted UV/optical continuum as for BL~Lac
objects. However, this scenario is not likely for most WLQs.
Shemmer et~al. (2009) found that the X-ray properties of
high-redshift WLQs are inconsistent with those of BL~Lac objects.
Furthermore, there is no evidence of strong optical variability or
polarization for these WLQs (see DS09; Meusinger et~al. 2011). The
UV-to-infrared SEDs of high-redshift WLQs are also similar to
those of typical quasars, while the SEDs of BL~Lac objects are
much different (DS09; Lane et~al. 2011). Nevertheless, it is
possible that the population of BL~Lac objects has a small
radio-quiet tail (e.g., Plotkin et~al. 2010b) and that a small
fraction ($\lesssim5\%$; see Lane et~al. 2011) of the general WLQ
population may be BL~Lac objects.

Most previous studies of WLQs were based on high-redshift 
objects. To investigate the nature of the overall
WLQ population, we obtained new \xray\ observations of
\hbox{low-redshift} WLQs selected mainly from the catalog of
radio-quiet BL~Lac candidates in Plotkin et~al. (2010a). We also
utilized sensitive archival \xray\ coverage of the sources in
their catalog. Our closely related science goals are the
following: (1) enable comparison of the broad-band SEDs of
\hbox{low-redshift} WLQs to those of high-redshift WLQs, typical
radio-quiet quasars, and BL~Lac objects; (2) provide basic
constraints upon \hbox{X-ray} spectral properties via band-ratio
analysis and joint spectral fitting; (3) clarify if there is
broad-band SED diversity among \hbox{low-redshift} WLQs; and (4)
allow reliable planning of future long, spectroscopic \hbox{X-ray}
observations.

In \S\ref{sample} we describe the selection of our sample of \hbox{low-redshift},
radio-quiet WLQ candidates.
In \S\ref{uvo} we detail their UV/optical observations and the measurement of
their rest-frame UV spectral properties.
In \S\ref{xray} we describe the relevant \hbox{X-ray} data analyses. Overall
results and associated discussion are presented in \S\ref{discuss}.
Throughout this paper, we adopt a cosmology with
$H_0=70.5$~km~s$^{-1}$~Mpc$^{-1}$,
$\Omega_{\rm M}=0.274$, and
$\Omega_{\Lambda}=0.726$
(e.g., Komatsu et~al. 2009).

% -----------------------------------------------------------------------------
% Sample Selection
% -----------------------------------------------------------------------------

%\section{Sample Selection}\label{sample}

\section{Selection of the \hbox{Low-Redshift} WLQ Candidates}\label{sample}

We obtained \chandra\ snapshot observations (3.0--4.1~ks) of six \hbox{low-redshift}
\hbox{($z=0.40$--$1.67$)} WLQ candidates. Five of the six targets were identified by
Plotkin et~al. (2010a) as radio-quiet, weak-featured SDSS quasars with all emission
features having REW~$\lesssim 5$~\AA. An additional source, SDSS~J0945+1009, was
similarly identified as a weak-featured quasar by Hryniewicz et~al. (2010). All the
objects are sufficiently bright in the optical band ($m_i\lesssim 18$) for short
\chandra\ observations to provide tight constraints on their X-ray-to-optical SEDs.

We further utilized the weak-featured quasar catalogs in Plotkin
et~al. (2010a) to search for \hbox{low-redshift}, radio-quiet
sources having sensitive archival \xray\ coverage. To ensure our
sample has the high \xray\ detection fraction necessary to provide
physically meaningful constraints, we only selected sources
covered by \chandra\ or \xmm\ observations.\footnote{We also
checked for pointed \rosat\ PSPC observations with an exposure
time greater than 5~ks and an off-axis angle less than $19'$
(within the inner ring of the PSPC detector). However, none of the
radio-quiet, \hbox{low-redshift} sources in the catalogs of
Plotkin et al. (2010a) is covered by \rosat\ observations meeting
these criteria.} An additional five sources were thereby added into
our sample. Three of them (J1013+4927, J1139$-$0201, and
J1604+4326) appear in the radio-quiet, weak-featured quasar
catalog (Table~6 in Plotkin et~al. 2010a). J1139$-$0201 was
targeted by \chandra\ as an optically selected BL~Lac candidate in
Cycle~5, while J1013+4927 and J1604+4326 were serendipitously
covered by \chandra\ or \xmm\ observations. The other two
objects (J2115+0001 and J2324+1443) were initially
identified as weak-featured quasars by Collinge et~al. (2005).
They were also listed in the catalog of Plotkin 
et~al. (2010a). These two sources did not have
constraints on their radio fluxes in Collinge et~al. (2005) or
Plotkin et~al. (2010a) but were later confirmed as radio-quiet
sources by the VLA observations of Plotkin et~al. (2010b).
They were targeted by \chandra\ as radio-quiet BL~Lac
candidates in Cycle~10; their observations were briefly reported in
Plotkin et~al. (2010b). Table~\ref{log_table} presents the \xray\ 
observation log for our sample. 

\begin{figure}[t]
    \centering
    \includegraphics[width=3.5in]{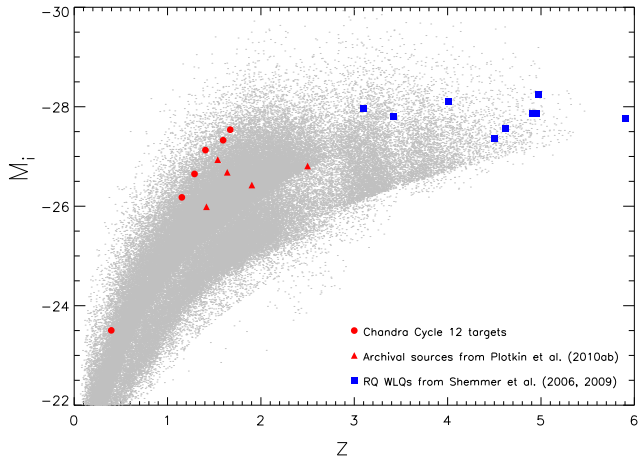}
    \caption{\footnotesize{SDSS absolute $i$-band magnitude, $M_i$, plotted versus redshift, $z$. The red 
              filled circles and triangles
              show our sample of \hbox{low-redshift} WLQ candidates; the blue filled 
              squares show high-redshift WLQs from Shemmer et al. (2006, 2009); the grey
              dots represent the 105,783 objects in the SDSS DR7 quasar catalog (Schneider et~al. 2010).}
             \label{zMi_fig}}
\end{figure}

\begin{figure*}[t]
    \centering
    \includegraphics[width=6.5in]{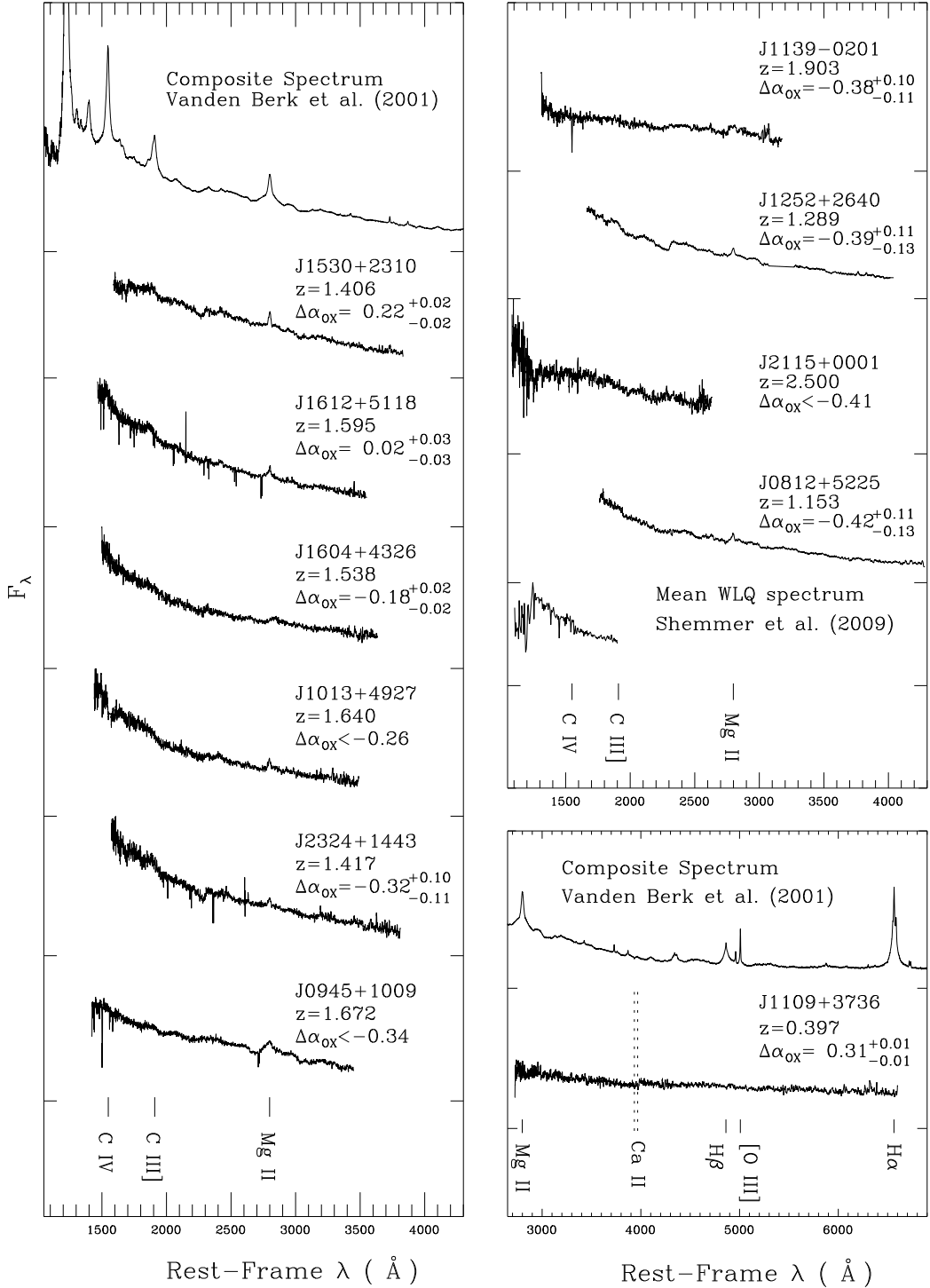}
    \caption{\footnotesize{SDSS spectra for the 11 sources in our sample of \hbox{low-redshift} WLQ candidates, 
             ordered by \daox\ (see \S\ref{xray} for definition). The \daox\ values and their 
             error bars (if the source is detected in \hbox{X-rays}) are shown for each source. 
             The name of each source is labeled in the format of 'J{\it hhmm+ddmm}'.
             The \hbox{$y$-coordinates} are the flux density ($F_\lambda$) in arbitrary linear units. 
             The tick marks on the \hbox{$y$-axis} show the zero flux-density level for each normalized
             spectrum. The spectra have been smoothed using a \hbox{5-pixel} \hbox{sliding-box} filter. 
             The spectrum of the radio-quiet BL~Lac candidate J1109+3736 is shown separately in the lower
             right panel for convenience of presentation (since its redshift is much lower than those of the
             other sources in our sample). Emission lines, including 
             \civ~$\lambda$1549, C~{\sc iii}]~$\lambda$1909, and Mg~{\sc ii}~$\lambda$2799, are labeled 
             in the left and upper right panels. The Mg~{\sc ii}~$\lambda$2799, H$\beta$~$\lambda$4862, 
             [O~{\sc iii}]~$\lambda$5007, and H$\alpha$~$\lambda$6564 lines are labeled in the lower right 
             panel; the Ca~{\sc ii} H/K break is also marked by the dotted lines. All the quoted values here 
             are vacuum wavelengths. The spectral resolution is 
             $R\approx$~2000. Also included are the composite spectrum 
             of SDSS quasars by Vanden~Berk et~al. (2001) and the mean spectrum of the high-redshift 
             WLQs of Shemmer et~al. (2009). }
             \label{spec_fig}}
\end{figure*}

Our sample includes 11 WLQs in total. All of the sources in our
sample have redshifts of $z\;<\;2.2$, except J2115+0001 which has a 
slightly higher redshift of $z=2.4995$ (see \S\ref{uvo:line} for 
redshift measurements). For comparison, all the
radio-quiet WLQs studied in \hbox{X-rays} by Shemmer et al. (2006,
2009) have $z>2.7$ (see Fig.~\ref{zMi_fig}). Fig.~\ref{spec_fig}
shows the SDSS spectra of the sources in our sample. The spectra
show no evidence for dust reddening or intrinsic broad absorption
lines (BALs); i.e., there is no indication that their UV/optical
continua or BELRs are obscured. We will compare the
multiwavelength properties of our sample to those of the
high-redshift WLQs in Shemmer et al. (2006, 2009) in
\S~\ref{discuss}.
% Figure 1: luminosity-redshift diagram

% Figure 2: source spectra

% -----------------------------------------------------------------------------
% UV/optical observations
% ----------------------------------------------------------------------------

\section{UV/Optical Observations}\label{uvo}

\subsection{UV Emission-Line Measurements}\label{uvo:line}

The redshift values (see Table~\ref{log_table}) for our \hbox{low-redshift} WLQs are
generally those from Hewett \& Wild (2010) which are the best available
measurements for large SDSS quasar samples. There are three sources lacking
Hewett \& Wild measurements. For two quasars (J1109$+$3736 and
J1139$-$0201), the redshift values are taken from the catalog of Plotkin et~al.
(2010a). The redshift of the other source (J2115+0001; $z=2.4995 \pm 0.0052$) is measured
based on a Ly$\alpha$ + \civ\ absorption system.\footnote{Plotkin et~al. (2010b) did 
not report the redshift for this source. In this work we adopt the redshift of the 
\hbox{Ly$\alpha$ + \civ} narrow absorption system as the systemic redshift. 
Nestor et~al. (2008) fit a Gaussian distribution centered at $v=0$~km~s$^{-1}$ with 
$\sigma=450$~km~s$^{-1}$ to the distribution of narrow \civ\ systems around quasar systemic
redshifts. We measured the redshift using that Gaussian dispersion
as the redshift uncertainty to obtain $ z = 2.4995 \pm 0.0052 $.}

To obtain accurate measurements of the weak emission lines, we manually measured
rest-frame emission-line properties for \civ, \ion{Si}{4},
the $\lambda 1900$ complex,\footnote{Mainly C~{\sc iii}]~$\lambda$1909, but also
including other features; see Note (b) of Table~\ref{qso_table}.} and
\ion{Fe}{3} UV48 (see Table~\ref{qso_table}) following the method in \S2.2 of Wu
et~al. (2011), which is summarized below. We first smoothed the SDSS spectra
with a 5-pixel sliding-box filter, and manually interpolated over strong narrow
absorption regions. We then fitted a power-law local continuum for each line
between their lower and upper wavelength limits $\lambda_{\rm lo}$ and $\lambda_{\rm hi}$
(see Table~2 of Vanden Berk et~al. 2001). After subtracting the local continuum,
we measured the REW value for each line. The \civ\ blueshifts were
calculated between the lab wavelength in the quasar rest frame ($1549.06$~\AA,
see Table~2 of Vanden Berk et~al. 2001) and the observed mode of all pixels with
heights greater than 50\% of the peak height, where mode =
3$\times$median$-2\times$mean. For comparison, we also include in
Table~\ref{qso_table} the corresponding measurements of the spectrum of \phl\
(Leighly et~al. 2007b) and of the composite spectrum of typical SDSS quasars in
Vanden~Berk et~al. (2001). The spectral measurements of \phl\ are included here
because some of our \hbox{low-redshift} WLQ candidates show similar unusual UV/optical
spectral properties to those of \phl\ (see \S\ref{discuss:class}).  The \mgii\
measurements from Shen et~al. (2011) are also listed in Table~\ref{qso_table}. These
measurements are reliable because the Fe~{\sc ii} component, which could affect
the \mgii\ strength measurement, was well modeled. These REW(\mgii) values somewhat
exceed the
selection criterion of REW~$\lesssim 5$~\AA\ for BL~Lac candidates in Plotkin
et~al. (2010a). This discrepancy mainly originates from differences in
measurement methods. For Plotkin et~al. (2010a), it was impractical to define
reference wavelengths to model the continuum in a uniform way for the entire
large sample since many objects lack redshift measurements. The REW values
in Plotkin et~al. (2010a) were measured manually after defining the continuum 
by eye for most sources.
While this method generally performed well for BL~Lac objects, it did not properly
model blended Fe emission for unbeamed objects.
%The \mgii\ lines of our objects
%are still notably weak compared to those of typical SDSS quasars.

Only two sources (J0945+1009 and J1612+5118) have high-quality
\civ\ coverage in their SDSS spectra so that we are able to
measure their \civ\ REW and blueshift values. Both sources have
weak and highly blueshifted \civ\ lines. J1139$-$0021 has no
clearly detectable \civ\ line in its SDSS spectrum; we could only
obtain an upper limit upon its REW. Therefore, we obtained \hbox{follow-up} UV spectroscopy
for this source with the Low-Resolution Spectrograph (LRS; 
Hill et~al. 1998) on the Hobby-Eberly Telescope (HET; 
Ramsey et~al. 1998). The UV emission-line measurements based on 
the HET spectroscopy are also listed in Table~\ref{qso_table}. 
J1139$-$0021 has a weak and strongly blueshifted \civ\ line in 
its HET spectrum. 

All of the sources
having \ion{C}{3}] coverage show weaker \ion{C}{3}] semi-forbidden
lines than those of typical quasars. The \ion{Fe}{3} UV48 strength
of our \hbox{low-redshift} WLQ candidates is generally similar to those of
typical quasars. The SDSS spectrum of J1109+3736 does
not have coverage of these rest-frame UV emission lines because of
its much lower redshift, while the signal-to-noise ratio of the
SDSS spectrum of J2115+0001 is too low to make reliable
emission-line measurements. 

% Figure 3: Line REW distribution
\begin{figure*}[t]
    \centering
    \includegraphics[width=6.0in]{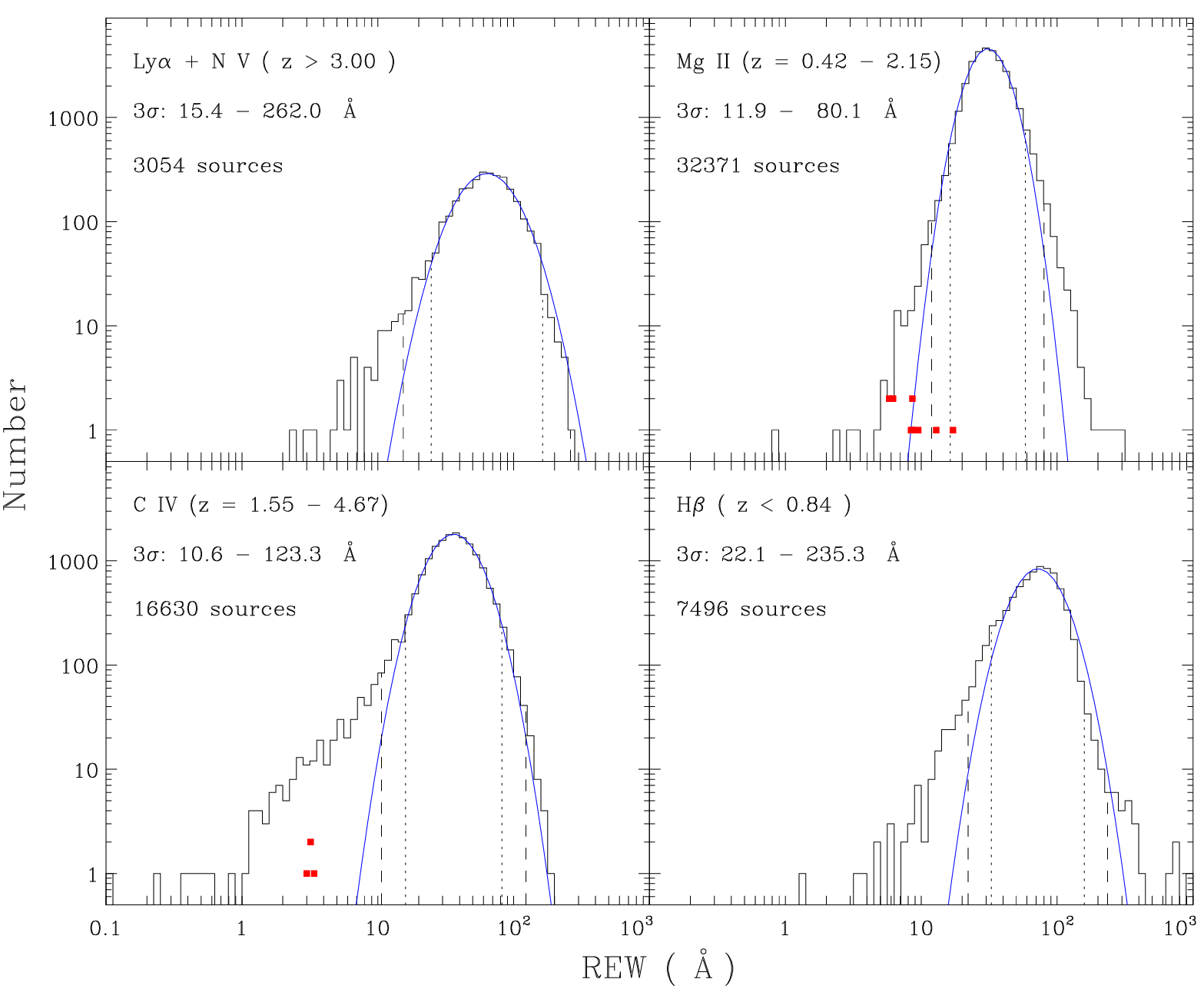}

    \vspace{0.1cm}
    \caption{\footnotesize{Distributions of REWs of \lyanv, \civ, \mgii, 
             and H$\beta$ for SDSS DR7 quasar samples described in \S\ref{uvo:criteria}. 
             In each panel, the blue solid line shows the best-fit lognormal 
             distribution. The dotted and dashed lines show the 2$\sigma$ and 3$\sigma$ ranges, 
             respectively. The 3$\sigma$ range is also noted in the upper right corner of each panel.
             The red filled squares with arbitrary $y$-coordinates are the REW values for our current sample. 
             The redshift range and the number of sources for each line REW 
             distribution are also noted. }
             \label{linehist_fig}}
\end{figure*}

\subsection{Comparing the Emission-Line Strengths of Low-Redshift WLQ Candidates and Typical SDSS Quasars}\label{uvo:criteria}

After measuring the strengths of the UV emission lines of our
low-redshift WLQ candidates (see Table~\ref{qso_table}), we
further investigated the REW distributions of prominent emission
lines that are covered by the SDSS spectra of \hbox{low-redshift}
quasars (such as \civ, \mgii, and H$\beta$). This allowed assessment 
of the emission-line weakness of our low-redshift, radio-quiet WLQ
candidates compared to typical SDSS quasars. Furthermore, our 
method of selecting low-redshift WLQ candidates is different from
that for high-redshift WLQs because \hbox{low-redshift} quasars do
not have coverage of \lyanv\ emission in their SDSS spectra. The
REW distributions of emission lines like \civ, \mgii, or H$\beta$
may give insights on universal selection criteria for WLQ
candidates at different redshifts.

We utilized the REW measurements of \civ, \mgii, and H$\beta$ from
the catalog of Shen et~al. (2011) for SDSS DR7 quasars (see their
Table~1). For the \civ\ and \mgii\ lines, we adopted the REW
values for the entire lines, while for H$\beta$ we added the REW
values of the broad and narrow components to obtain the REW of the
entire line. Shen et~al. (2011) reported the REW distributions of
these lines for all SDSS DR7 quasars having applicable REW measurements
(see their Figs.~12--14). In this work, we selected 
unbiased samples of SDSS quasars with high-quality optical/UV spectra
to study the REW distributions of these emission lines by imposing 
the following criteria: 
\begin{enumerate}
\item We only 
use the DR7 quasars selected with the final algorithm given by 
Richards et~al. (2002) to maintain consistency with DS09. 
\item BAL 
quasars that were cataloged in Gibson et~al. (2009) and Shen 
et~al. (2011) were removed. 
\item We restricted the redshift ranges 
of the objects for each line (\civ: $z=1.55$--$4.67$; \mgii: 
$z=0.42$--$2.15$; H$\beta$: $z<0.84$) to ensure that the SDSS 
spectra of these objects cover the whole region of each 
emission line defined by $\lambda_{\rm lo}$ and $\lambda_{\rm hi}$ 
listed in Table~2 of Vanden~Berk et~al. (2001). 
\item To select
objects with high-quality SDSS spectra, we calculated the 
signal-to-noise ratio ($S/N$) of continuum regions close
to each emission line, $SN_{1700}$ for \civ, $SN_{3000}$ for \mgii,
and $SN_{5150}$ for H$\beta$ (see Table~\ref{contsn_table}), following the 
method of Gibson et~al. (2009). $SN_{1700}$, $SN_{3000}$, and $SN_{5150}$
are calculated as the median of the ratio between the flux and the 
error (obtained from the SDSS pipeline) for all the spectral bins 
in the rest-frame 1650--1750~\AA, 2950--3050~\AA, and 5100--5200~\AA\ regions, 
respectively. These wavelength regions are free of strong emission
and/or absorption features, but are still close to the above emission
lines in our study. We require each of the above continuum $S/N$
to be greater than 7. The $S/N$ values of the emission lines themselves
were not utilized because that would introduce bias against objects
with weak emission lines. 
\item We eliminated the objects that have 
large fractions of bad pixels in their SDSS spectra in the wavelength ranges of 
these emission lines (these lead to unreliable measurements). 
We imposed the following cuts on the numbers
of pixels that were included in the fitting for each line given 
in the catalog of Shen et~al. (2011): \verb+LINE_NPIX_CIV+ $> 250$
for \civ, \verb+LINE_NPIX_MGII+ $> 300$ for \mgii, and \verb+LINE_NPIX_HB+ 
$> 150$ for H$\beta$. 
\end{enumerate}
Applying the above cuts on redshift, continuum 
$S/N$, and the numbers of pixels in the fitting (criteria 3--5 above) 
removed 35\%, 27\%, and 
37\% of the objects in the REW distribution investigations 
for \civ, \mgii, and H$\beta$, respectively. 
All the above restrictions and quality cuts
are necessary because unreliable line measurements could 
significantly affect the REW distributions particularly in the tails
with low or high REW values.

Fig.~\ref{linehist_fig} shows the REW distributions of the \civ,
\mgii, and H$\beta$ lines for our selected samples of 
SDSS quasars. We also
include the histogram for \lyanv\ from DS09 for comparison. DS09
also showed a histogram for \civ\ REWs of SDSS DR5 quasars, which
has a similar profile as the \civ\ REW histogram in
Fig.~\ref{linehist_fig}. We fit the REW histogram of each line
with a lognormal distribution (see the blue solid lines in
Fig.~\ref{linehist_fig}; also see the dotted and dashed lines for
the $2\sigma$ and $3\sigma$ ranges of each lognormal model). The
REW measurements for the quasars in our \hbox{low-redshift} WLQ
sample (see \S\ref{uvo:line}) are also shown in
Fig.~\ref{linehist_fig}. All the \civ\ REW values of our sources
are far below the negative 3$\sigma$ deviation of the lognormal
distribution. Most of the REW(\mgii) values are also below the
negative 3$\sigma$ deviation of the lognormal distribution; the
largest REW(\mgii) value for our sample is close to the negative
2$\sigma$ deviation (see the top right panel of
Fig.~\ref{linehist_fig}).

\begin{figure}[t]
    \centering
    \includegraphics[width=3.6in]{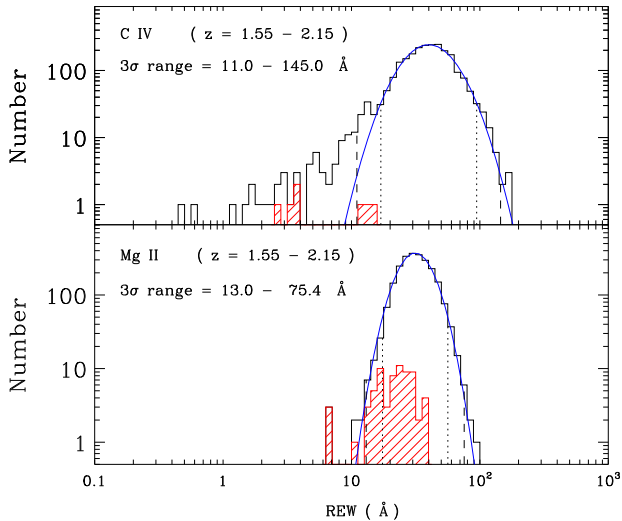}
    \caption{\footnotesize{Distributions of REWs for \civ\ (top panel) and \mgii\ (bottom panel)
             for the same set of SDSS quasars with $1.55\;<\;z\;<\;2.15$ and 
             $45\;<\;\log L_{3000}\;<\;46$. The red shaded histogram in the top panel shows 
             the REW(\civ) distribution for sources with REW(\mgii) below the 3$\sigma$ negative 
             deviation for \mgii. The red shaded histogram in the bottom panel shows the REW(\mgii) 
             distribution for sources with REW(\civ) below the 3$\sigma$ negative deviation for \civ.
             The bottom panel shows that many objects with weak \civ\ emission do not have 
             weak \mgii. Other lines follow the same definitions as those in Fig.~\ref{linehist_fig}.}
             \label{linehistcomp_fig}}
\end{figure}

The REW distribution of \lyanv\ in DS09 shows a prominent tail
toward low REW values; this tail is the basis on which the
high-redshift WLQs are defined. The \civ\ REW distribution shows
similar behavior in that a more prominent skew tail toward low REW values
exists, while there is no corresponding tail toward high REW
values. However, the histogram of the \mgii\ REWs appears
more symmetric, with only small tails toward both the low end 
and high end of the REW distribution. The H$\beta$ REW
distribution is similar to that for \lyanv; a prominent tail
toward low REW values exists, while the tail toward high REW 
values is much less significant. We randomly chose sets of sources in 
the tails with low or high REW values in the histograms for 
\civ, \mgii, and H$\beta$ and then visually examined their SDSS spectra 
and the quality assessment 
plots\footnote{See https://www.cfa.harvard.edu/~yshen/BH\_mass/dr7.htm.} 
for their individual spectral fits in Shen et~al. (2011). These
sources have good spectral-fit quality; their REW measurements should
be reliable. It is worth noting that the quasars
in the REW histograms for various lines have different ranges of
redshift and luminosity, which may affect their REW distributions
(e.g., via the Baldwin effect; Baldwin 1977). We test this
hypothesis by comparing the \civ\ and \mgii\ REW distributions for
a set of SDSS quasars with $1.55\;<\;z\;<\;2.15$ and $45\;<\;\log
\nu L_{3000}\;<\;46$ ($\nu L_{3000}$ is the luminosity at
rest-frame 3000~\AA\ in erg~s$^{-1}$, obtained from Shen et~al.
2011). Fig.~\ref{linehistcomp_fig} shows that both the \civ\ and
\mgii\ REW distributions show similar behavior to the
distributions presented in Fig~\ref{linehist_fig}; the \civ\ 
histogram has a significant tail at the low REW end, while the 
\mgii\ histogram is symmetric with weak tails. 
While the physical reason for the different emission-line REW
distributions is unclear, the \mgii\ line has low
optical depth, while the \civ\ line has much
higher optical depth (e.g., see Eracleous et~al. 2009). 
In the context of a disk-wind model for the BELR (e.g., Murray \& Chiang 
1997; Proga et~al. 2000), the \civ\ emission is considered to be 
mainly from the AGN wind, while the \mgii\ line mostly originates
from the accretion disk (e.g., Leighly 2004; Richards et~al. 2011).
%The \civ\ emission is considered to be mainly from the AGN wind, 
%while the \mgii\ line mostly originates from the accretion disk 
%(e.g., Murray \& Chiang 1997; Elvis 2000).

% Figure 4: Line REW comparison

% Figure 5: Line REW correlation
\begin{figure*}[t]
    \centering
    \includegraphics[width=3.0in]{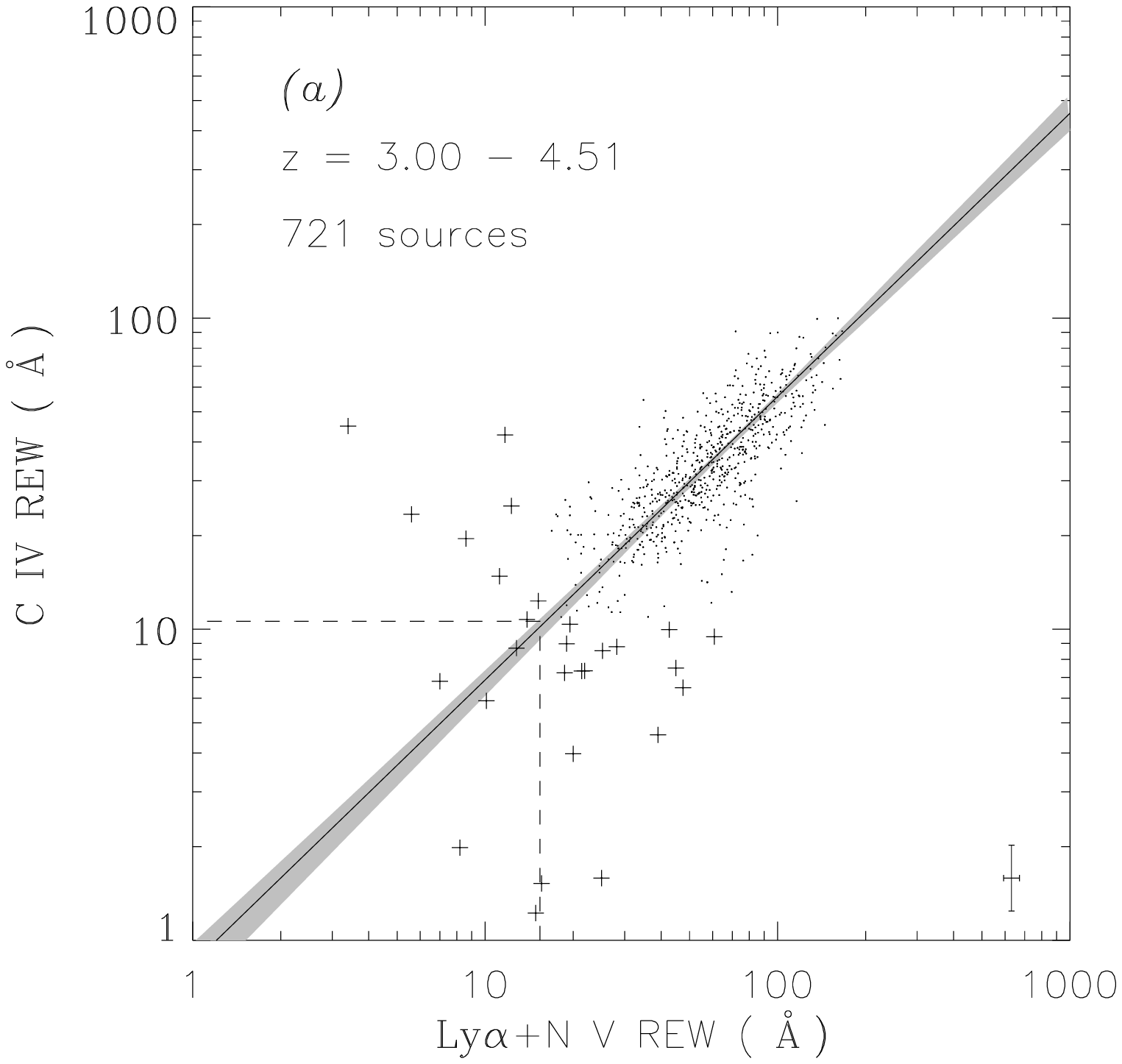}
    \includegraphics[width=3.0in]{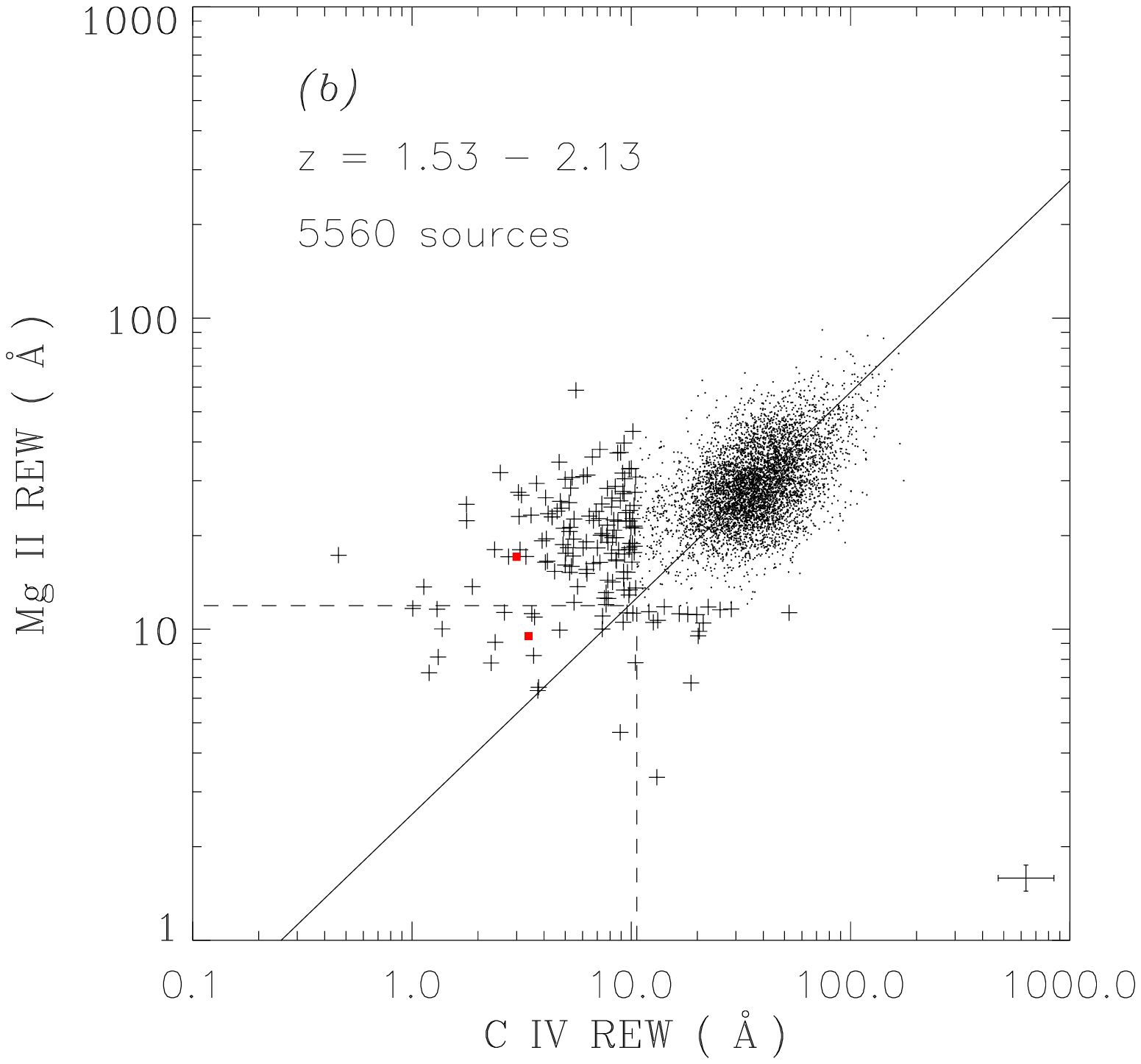}
    \includegraphics[width=3.0in]{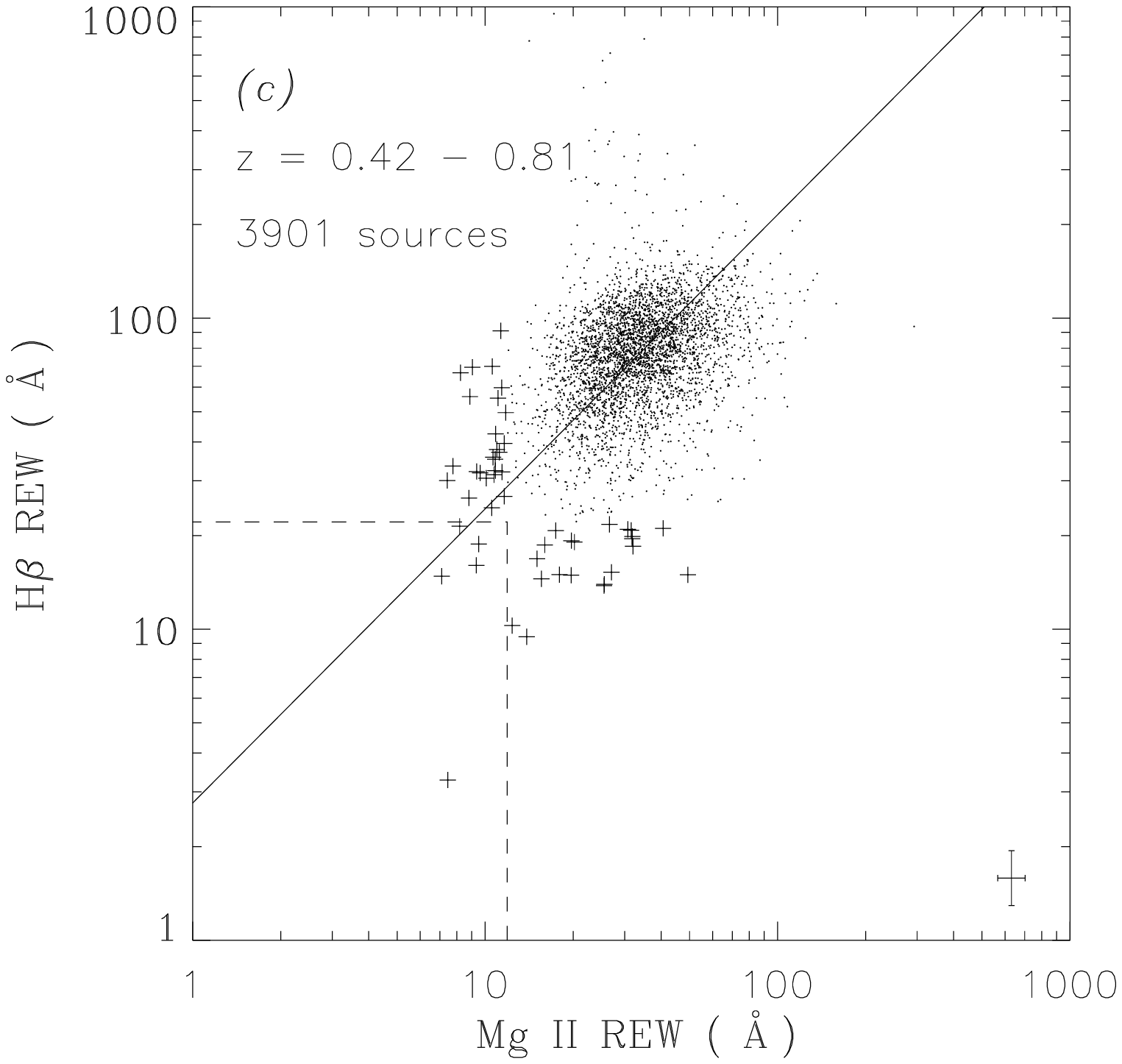}
    \caption{\footnotesize{Correlations between line REWs: $(a)$ Ly$\alpha$~vs.~\civ; $(b)$ 
             \civ~vs.~Mg~{\sc ii}; $(c)$ Mg~{\sc ii}~vs.~H$\beta$. No upper limits are present for any 
             REW measurements. The median errors of the REW measurements for sources with the corresponding  
             weak emission line (i.e., REW values below the $3\sigma$ negative deviation) are shown in 
             the lower right corner of each panel. The solid lines show the best-fit 
             power-law models. The dashed lines show the 3$\sigma$ values of the weak REW tails in 
             Fig.~\ref{linehist_fig}. The red filled squares show the sources in our current sample. 
             The crosses in each panel are the sources having REW values below the negative 3$\sigma$
             deviation for either of the emission lines in the panel. The grey-shaded area in panel 
             $(a)$ shows the 90\% confidence uncertainty range of the best-fit correlation. 
             The redshift range and the total number of sources in each panel are also noted. }
             \label{linecorr_fig}}
\end{figure*}

The $3\sigma$ tail toward low REW values of the \civ\ histogram is
defined by REW(\civ)$\;\lesssim\;10$~\AA, which is consistent with
that derived in DS09. Using this criterion one could perhaps hope to 
extend the redshift range of WLQ selection from $z\;>3$ down to
$z\;>1.5$. However, this definition of WLQs has significant
inconsistency with that in DS09 based on REW(\lyanv)$\;<\;15.4$~\AA. 
Fig.~\ref{linecorr_fig}$(a)$ shows the relation between the REW
values of \lyanv\ and \civ, and the best-fit power-law model found
using the IDL routine {\sc linmix\_err}.\footnote{The {\sc
linmix\_err} procedure is a Bayesian approach to linear regression
which usually has good performance when there is significant
intrinsic scatter and correlated error bars (see Kelly 2007 for
more details). To identify the best-fit power-law model, we first
fit the correlation by assigning REW(\lyanv) as the
``independent'' variable and REW(\civ) as the ``dependent''
variable, and then exchange these two variables to obtain another
fitting correlation. We finally calculated the bisector of these
two power-law models as the best-fit model (e.g., Isobe et~al. 1990). We
used the same bisector method for the correlations between the REW
values of other emission lines.} The shaded area shows the 90\%
confidence uncertainty range obtained via a nonparametric
bootstrap method (Efron 1979). Although the REW values of these
two lines are positively correlated, large scatter exists. Over
half of the sources with REW(\lyanv)$\;<\;15.4$~\AA\ have
REW(\civ)$\;>\;10$~\AA, and vice versa (see Fig.~\ref{linecorr_fig}$a$).
The large scatter is likely to be intrinsic, since the REW measurement errors (see
Fig.~\ref{linecorr_fig} for typical error bars) are much smaller
than the scatter. DS09 also suggested the inconsistency between 
sources with weak \lyanv\ and those with weak \civ\ emission. A
total of 39 of the 74 WLQs cataloged in DS09 have 
REW(\civ)$\;>\;10$~\AA\ based on the measurements in Shen et~al. 
(2011).\footnote{It is worth noting that many WLQs in the DS09 sample have
large uncertainties on their \civ\ measurements. Only 13 of the 74 WLQs
definitely have REW(\civ)$\;>\;10$~\AA\ at $3\sigma$ significance.} 
As stated in DS09, their measurements may underestimate
REW(\lyanv) when strong intervening absorption exists; this is
perhaps one reason for the inconsistency. 
%DS09 flagged
%their WLQs with REW(\civ)$\;>\;10$~\AA\ (see their Table~2; based on
%the measurements in Shen et~al. 2008) and those with proximate
%damped Ly$\alpha$ absorbers (PDLAs; e.g., Prochaska et~al. 2008).
%They did analysis on both their full WLQ sample and a subset 
%excluding the sources having REW(\civ)$\;>\;10$~\AA\ or PDLAs, and 
%obtained similar results.  
The correlation between
the REWs of \civ\ and \mgii\ also has significant scatter, and so
does the correlation between the REWs of \mgii\ and H$\beta$ (see
Figs.~\ref{linecorr_fig}$b$ and \ref{linecorr_fig}$c$). The red shaded
histograms in Fig.~\ref{linehistcomp_fig} show the REW(\civ) distribution
for sources with weak \mgii\ and the REW(\mgii) distribution for 
sources with weak \civ. While the sources with weak \mgii\ also tend to 
have weak \civ\ (below the negative $2\sigma$ deviation), many objects
with weak \civ\ have fairly strong \mgii\ (see the bottom panel of Fig.~\ref{linehistcomp_fig}).  

Given the results above, it is therefore difficult to find consistent criteria for WLQs at
different redshifts using different emission lines even though the REW 
distributions of \civ\ and H$\beta$ show similar behavior to that of \lyanv. 
Since there is
no single line that is covered by SDSS spectroscopy for quasars at
all redshifts between zero and six, it appears that there is not a  
straightforward, direct way to define universal selection criteria for WLQs 
at all redshifts solely based on SDSS spectroscopy. Therefore, we only choose our
\hbox{low-redshift} WLQ candidates mainly from the catalog of
Plotkin et~al. (2010a), which has a strict criterion on all 
emission-line strengths (REW~$\lesssim 5$~\AA). Future UV
spectroscopy which covers the \lyanv\ and/or the \civ\ regions for
\hbox{low-redshift} WLQs will provide insights toward
a universal definition for WLQs.

% -----------------------------------------------------------------------------
% X-Ray Data Analysis
% -----------------------------------------------------------------------------

\section{X-ray Data Analysis}\label{xray}

The six new \hbox{low-redshift} WLQ candidates targeted in
\chandra\ Cycle~12 were observed with the S3 CCD of the Advanced
CCD Imaging Spectrometer (ACIS; Garmire et~al. 2003). The
reduction of the \chandra\ data was performed using standard CIAO
v4.3 routines. \xray\ images were produced for the observed-frame
soft \hbox{(0.5--2.0 keV)}, hard \hbox{(2.0--8.0 keV)}, and full
\hbox{(0.5--8.0 keV)} bands using {\it ASCA} grade 0, 2, 3, 4, and
6 events. The {\sc wavdetect} algorithm (Freeman et~al. 2002) was
run on the images using a detection threshold of $10^{-5}$ and
wavelet scales of $1$, $\sqrt{2}$, $2$, $2\sqrt{2}$, and $4$
pixels. All targets, except J0945$+$1009, were detected by \chandra\
within $0.8''$ of the optical coordinates. The \xray\ images of 
J0945$+$1009 were visually examined, and no hint of a detection was found. 
Aperture photometry was
performed using the IDL {\sc aper} procedure on each object. An
aperture radius of $1.5''$ was adopted for each source
($\approx95\%$ enclosed energy for soft band, $\approx90\%$
enclosed energy for hard band; aperture corrections were applied)
except J1109$+$3736 and J1530$+$2310, for which the aperture
radius was $3.0''$ because of their large numbers of detected
\xray\ counts (see Table~\ref{cts_table}; no aperture corrections
were applied to these two sources). The background region for each
source was defined as an annulus with inner and outer radii of
twice and three times the aperture radius. All background
regions are free of \xray\ sources. The upper limits upon \xray\
counts for J0945+1009 were determined using the method of Kraft et
al. (1991) at 95\% confidence. Table~\ref{cts_table} lists the
\xray\ counts in the three bands, as well as the band ratio
(defined as the ratio between hard-band counts and soft-band
counts) and effective \pl\ photon index for each source. The
effective \pl\ photon index was determined from the band ratio
using the \chandra\
PIMMS\footnote{http://cxc.harvard.edu/toolkit/pimms.jsp} tool,
under the assumption of a \pl\ model with Galactic absorption
only.

Four archival sources (J1139$-$0201, J1604$+$4326, J2115$+$0001,
and J2324$+$1443) were observed by \chandra\ in Cycles 5, 7, and
10. All of the quasars were detected by \chandra\ except
J2115$+$0001. Similar \chandra\ data-reduction and processing
procedures were performed for these objects. Three of them
(J1139$-$0201, J2115$+$0001, and J2324$+$1443) were targeted in
their \chandra\ observations, for which the aperture radius was
set to be $1.5''$. J1604$+$4326 was serendipitously covered by the
ACIS-I detector in two \chandra\ observations. We measured the
\xray\ counts individually for these two observations, and then
calculated the mean count rate and flux in the soft band. This source
did not show significant variability ($\lesssim 12\%$) between its two 
\chandra\ observations ($\approx10$ hours apart in the quasar rest frame). The
aperture radius ($2.6''$) for this source was determined to be the
95\% enclosed-energy radius at $1.497$~keV based on the point
spread function (PSF) of the ACIS detector at an off-axis angle of
$3.7'$. The \chandra\ observations of J2115$+$0001 and
J2324$+$1443 were briefly reported in Plotkin et~al. (2010b), and our
results are consistent with theirs.

One archival source (J1013$+$4927) was serendipitously covered by \xmm\ on 2004
April~23. Data reduction and processing were performed using standard \xmm\
Science Analysis System (v10.0.0) routines. This source is undetected by both
the \verb+MOS+ and \verb+pn+ detectors using the {\sc eboxdetect} procedure.
Visual inspection of the images verifies the \hbox{non-detection} of this source.
We only used the data from the \verb+MOS+ detectors because they have higher
angular resolution, which enables more reliable count extraction and
background estimation. The events files were filtered by removing background
flaring periods (12\% of the total exposure time) in which the count rate
exceeded $0.35$~s$^{-1}$ for events with energies above 10~keV. The aperture for
photometry ($49.9''$ radius) was taken to be the 90\% enclosed-energy radius
at $1.5$~keV based on the PSF of the \verb+MOS+ detectors at an off-axis angle
of $6.4'$. The upper limits upon \xray\ counts were determined to be $3\sqrt{N}$,
where $N$ is the total counts within the aperture.

Table~\ref{aox_table} lists the key \xray, optical, and radio properties of our \hbox{low-redshift}
WLQs:

Column (1): The SDSS equatorial coordinates (J2000) for the source.

%Column (2): The quasar's redshift. These values were taken from the SDSS Catalog
%;Archive Server (CAS), except for four sources having larger uncertainties (see \S2 for details).

%the improved
%measurements in Hewett \& Wild (2010), which have significantly reduced systematic
%biases compared to the redshift values in the SDSS DR5 quasar catalog.

Column (2): The apparent $i$-band magnitude of the source using the
SDSS quasar catalog BEST photometry.

Column (3): The absolute $i$-band magnitude for the
source, $M_{i}$, from the SDSS DR7 quasar catalog (Schneider
et~al. 2010), calculated by correcting for Galactic extinction and
assuming a \pl\ spectral index of $\alpha_\nu\;=\;-0.5$ (e.g.,
Vanden~Berk et~al. 2001).

Column (4): The Galactic neutral hydrogen column density in units of $10^{20}$~cm$^{-2}$,
obtained with the \chandra\ COLDEN\footnote{http://cxc.harvard.edu/toolkit/colden.jsp}
tool.

Column (5): The count rate in the observed-frame soft \xray\ band
($0.5$--$2.0$~keV), in units of $10^{-3}$~s$^{-1}$. For the two off-axis sources (J1013+4927 and
J1604+4326), the count rate (or upper limit) is corrected for vignetting using
exposure maps.

Column (6): The Galactic absorption-corrected flux in the observed-frame
soft \xray\ band in units of $10^{-14}$~erg~cm$^{-2}$~s$^{-1}$,
obtained with the \chandra\ PIMMS tool. An absorbed \pl\ model was utilized with a photon index $\Gamma=2$,
which is typical for quasars, and the Galactic neutral hydrogen column density for each source
($N_H$, given in Column~4).

Column (7): The Galactic absorption-corrected flux density at
rest-frame 2 keV in units of $10^{-32}$ erg cm$^{-2}$ s$^{-1}$ Hz$^{-1}$, obtained with
the \chandra\ PIMMS tool.

Column (8): The logarithm of the quasar \xray\ luminosity in the rest-frame
$2$--$10$~keV band corrected for Galactic absorption.

Column (9): The continuum flux density at rest-frame 2500~\AA~ in
units of $10^{-27}$ erg~cm$^{-2}$~s$^{-1}$~Hz$^{-1}$, from the SDSS quasar spectral property
catalog in Shen et~al. (2011).
%using a method
%similar to that described in \S2.2 of Vignali et~al.~(2003). It is calculated as an
%interpolation between the fluxes of two SDSS filters bracketing rest-frame 2500~\AA.

Column (10): The logarithm of the
monochromatic luminosity at rest-frame 2500~\AA, derived from the flux density at
rest-frame 2500~\AA. A cosmological bandpass correction is utilized.

Column (11): The X-ray-to-optical \pl\ slope, given by
\begin{equation}
    \alpha_{\rm ox} = \frac{{\rm log}(f_{\rm 2\;keV} /
    f_{2500\mbox{\rm~\scriptsize\AA}})}{{\rm log}(\nu_{\rm 2\;keV} / \nu_{2500\mbox{\rm~\scriptsize\AA}})}
    = 0.384\ {\rm log} \bigg(\frac{f_{\rm 2\;keV}}{f_{2500\mbox{\rm~\scriptsize\AA}}}\bigg) .
\label{aoxeqn}
\end{equation}
The flux density is measured per unit frequency.
%Note our UV/optical and \xray\ measurements were not simultaneous.

Column (12): \daox, a parameter assessing the relative \xray\ brightness
(see \S\ref{discuss:daox}), defined as
\begin{equation}
    \Delta\alpha_{\rm ox} = \alpha_{\rm ox(measured)} - \alpha_{\rm ox(expected)}.
\label{daoxeqn}
\end{equation}
The expected \aox\ for a typical radio-quiet quasar is calculated
using the $\alpha_{\rm ox}$-$L_{2500\mbox{\rm~\scriptsize\AA}}$ correlation given as
Equation (3) of Just et~al.~(2007). The statistical significance of \daox\
(given in parentheses) is in units of $\sigma$, which is
obtained from Table~5 of Steffen et~al.~(2006) as the RMS for \aox\ of quasars with
several ranges of luminosity.

Column (13): The factor of \xray\ weakness, derived from the \daox\ values in Column~(12),
quantifying the \xray\ weakness of our sources compared to a typical radio-quiet
quasar with similar UV/optical
luminosity, calculated as $f_{\rm x-weak}=10^{-\Delta\alpha_{\rm ox}/0.384}\approx403^{-\Delta\alpha_{\rm ox}}$.
A source with \daox$=-0.384$ has an \xray\ flux only $\approx10\%$ that of typical quasars,
corresponding to an \xray\ weakness factor of $\approx10$.

Column (14): The optical-to-radio \pl\ slope, given by
\begin{equation}
    \alpha_{\rm ro} = \frac{{\rm log}(f_{\rm 5\;GHz} / f_{2500\mbox{\rm~\scriptsize\AA}})}
     {{\rm log}(\nu_{\rm 5\;GHz} / \nu_{2500\mbox{\rm~\scriptsize\AA}})}.
%    = 0.384\ {\rm log} \bigg(\frac{f_{\rm 2\;keV}}{f_{2500\mbox{\rm~\scriptsize\AA}}}\bigg)
%    R = \frac{f_{\rm 5\;GHz}}{f_{4400\mbox{\rm~\scriptsize\AA}}}{\rm .}
\label{aroeqn}
\end{equation}
The values of $f_{2500\mbox{\rm~\scriptsize\AA}}$ are given in
Column~(9). The values of $f_{5\;{\rm GHz}}$ were calculated using
a radio \pl\ slope of $\alpha_\nu$ = $-0.8$ (e.g., Falcke et~al. 
1996; Barvainis et~al. 2005) and a flux at
an observed-frame wavelength of 20 cm, $f_{20\;{\rm cm}}$. For
sources detected by the {\it FIRST} survey (Faint Images of the
Radio Sky at Twenty-Centimeters; Becker et~al. 1995), $f_{20\;{\rm
cm}}$ was taken from the {\it FIRST} source catalog. For sources
covered but not detected by the {\it FIRST} survey, the upper
limits for radio flux density were placed as $f_{20\;{\rm
cm}}<0.25+(5\sigma_{\rm rms})$~mJy, where $\sigma_{\rm rms}$ is
the RMS noise of the {\it FIRST} survey at the object's
coordinates (see \S5.3.1 of Plotkin et~al. 2010a for more
details). For the two sources not covered by the {\it FIRST}
survey (J2115+0001 and J2324+1443), the radio flux density was
measured via targeted VLA observations (see \S3.1 of Plotkin
et~al. 2010b).

The \aro\ parameter is related to the commonly used radio-loudness
parameter, \hbox{$R=f_{5\;{\rm
GHz}}/f_{4400\mbox{\rm~\scriptsize\AA}}$} (e.g., Kellermann et~al.
1989), by the following equation:
\begin{equation}
\alpha_{\rm ro} = \frac{{\rm log}[R(2500/4400)^{\alpha_\nu}]}
     {{\rm log}(\nu_{\rm 5\;GHz} / \nu_{2500\mbox{\rm~\scriptsize\AA}})}
     = -0.186\;{\rm log}\;R -0.023,
\label{reqn}
\end{equation}
where we use $\alpha_\nu = -0.5$. Therefore, we have \aro$>-0.21$
for radio-quiet quasars ($R<10$); \hbox{$-0.39<\;$\aro$<-0.21$}
for radio-intermediate quasars ($10<R<100$); and \aro$<-0.39$ for
radio-loud quasars ($R>100$).  All the \hbox{low-redshift} WLQ
candidates in our sample are radio quiet. Note that the radio and
optical observations of our sources are non-simultaneous. While
the non-simultaneity does not generally alter the classification
of typical radio-quiet quasars, it may significantly affect that
of BL~Lac objects because of their rapid and large-amplitude
variability.

% Figure 6: aox histogram
\begin{figure*}[t]
    \centering
    \includegraphics[width=6.5in]{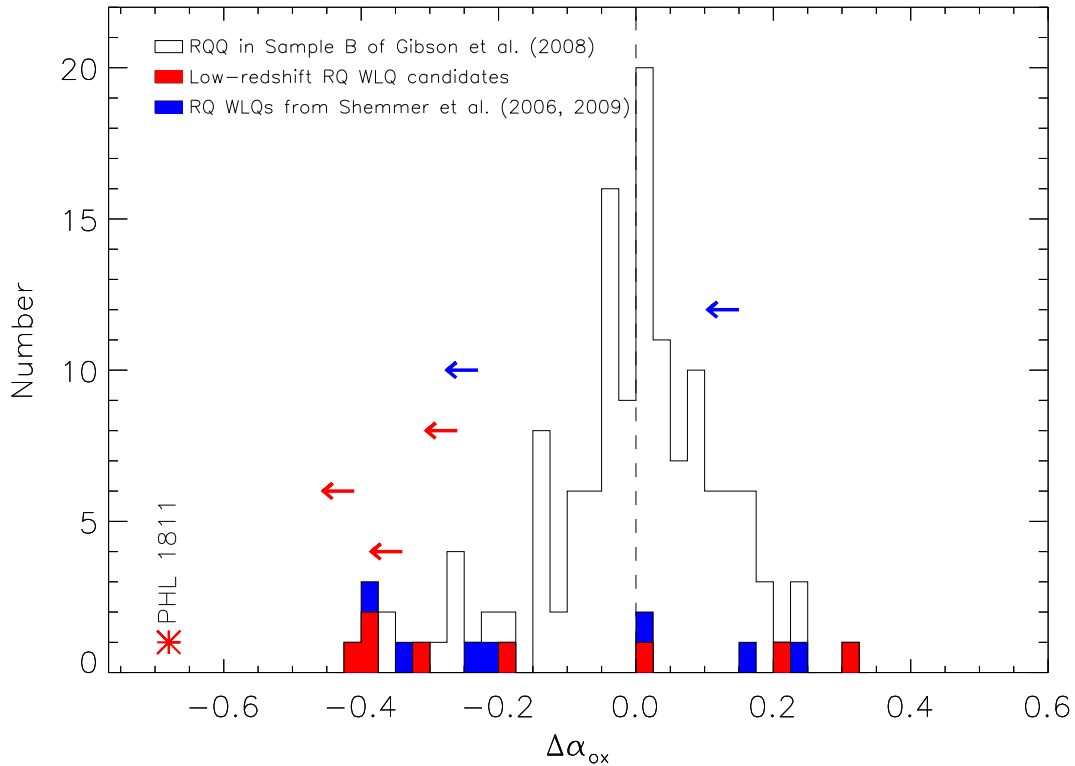}
    \caption{\footnotesize{Distribution of \daox\ values for the WLQ candidates, compared to that of the 132 
             \hbox{radio-quiet}, \hbox{non-BAL} quasars in Sample~B of Gibson et al. (2008). The red
             histograms and red leftward arrows represent the \hbox{low-redshift} sources in our sample which 
             are detected (8 sources) and undetected (3 sources) in \hbox{X-rays}, respectively. The blue 
             histogram and blue leftward arrows represent the high-redshift sources from Shemmer et al. 
             (2006, 2009) which are detected (7 sources) and undetected (2 sources) in \hbox{X-rays}, 
             respectively. The unshaded histogram shows the 
             \hbox{radio-quiet}, \hbox{non-BAL} quasars in Sample~B of Gibson et al. (2008). 
             PHL 1811 is shown as a red asterisk. The dashed vertical line shows \daox $=0$. Note the many 
             ($>50\%$) WLQs with \daox$\;<-0.2$.}
             \label{daox_fig}}
\end{figure*}

% -----------------------------------------------------------------------------
% Discussion
% -----------------------------------------------------------------------------

\section{Results and Discussion}\label{discuss}

\subsection{Relative X-ray Brightness}\label{discuss:daox}

The \daox\ parameter (see Eqn.~\ref{daoxeqn} for definition) is
utilized to assess the \xray\ brightness of a quasar relative to
typical radio-quiet quasars with similar UV luminosity. We compare
the \daox\ distribution of our \hbox{low-redshift}, radio-quiet
WLQ candidates (see Fig.~\ref{daox_fig}) to that of the 132
radio-quiet, non-BAL quasars in Sample B of Gibson et~al.
(2008),\footnote{We used an improved version of the Sample~B
quasars in Gibson et~al. (2008) from which we further removed
seven BAL quasars  (see Footnote~16 in Wu et~al. 2011).} 
which represent typical
radio-quiet SDSS quasars. All of the 132 Sample~B quasars are
\xray\ detected. The Peto-Prentice test (e.g., Latta 1981),
implemented in the Astronomy Survival Analysis package (ASURV;
e.g., Lavalley et~al. 1992), is used to assess whether our
\hbox{low-redshift} WLQ candidates follow the same \daox\
distribution as that for typical quasars (see results in
Table~\ref{twost_table}). We prefer the Peto-Prentice test to 
other possible similar tests because
it is the least affected by the factors of different censoring
patterns or unequal sizes of the two samples which exist in
our case. We also compare the \daox\ distribution of
high-redshift, radio-quiet WLQs in Shemmer et~al. (2006, 2009) to
that of our \hbox{low-redshift}, radio-quiet WLQ candidates and
that of typical SDSS quasars (also see Fig.~\ref{daox_fig} and
Table~\ref{twost_table}).

The \daox\ distribution of our \hbox{low-redshift}, radio-quiet
WLQ candidates is significantly different from that of typical
SDSS quasars. The probability of null-hypothesis (two samples
following the same distribution) is only $6.3\times 10^{-7}$. This
result is mainly due to the presence of a skew tail of \xray\ weak
WLQs (see Fig.~\ref{daox_fig}). Seven out of the 11 objects in our 
sample of \hbox{low-redshift}, radio-quiet WLQ candidates have 
\daox$\;<-0.2$, giving a fraction of \xray\ weak objects of 
($64^{+34}_{-24}$)\% (68\% confidence level). 
The mean \daox\ value for the \hbox{low-redshift},
radio-quiet WLQ candidates is $-0.214\pm0.078$, calculated with
the Kaplan-Meier estimator also implemented in the ASURV package,
while that for the Sample B quasars is $-0.001\pm0.011$. The
\daox\ distribution of the nine high-redshift, radio-quiet WLQs in
Shemmer et~al. (2006, 2009) is also different from that of typical
radio-quiet SDSS quasars, but less significantly (the probability
of null-hypothesis is $7.1\times 10^{-3}$).\footnote{This result
is somewhat inconsistent with the finding by Shemmer et~al. (2009)
that their high-redshift WLQs have a similar \daox\ distribution
to that of typical SDSS quasars. Shemmer et~al. (2009) used the
Kolmogorov-Smirnov test and ignored the two high-redshift,
radio-quiet WLQs with \daox\ upper limits. We include all nine
high-redshift, radio-quiet WLQs since the Peto-Prentice test can
properly treat censored data. The utilization of an improved Sample~B 
(see Footnote~17) does not substantially contribute to the inconsistency 
here. The Peto-Prentice test using all nine high-redshift, radio-quiet WLQs 
and the original Sample~B (as used by Shemmer et~al. 2009) provides a 
null-hypothesis probability of 
$1.53\times 10^{-2}$.} Five of the nine objects in the  
\hbox{high-redshift}, radio-quiet WLQ sample have 
\daox$\;<\;-0.2$, giving a fraction of \xray\ weak objects of 
($56^{+37}_{-24}$)\% (68\% confidence level); we note that the fraction could 
be somewhat higher (6/9) owing to the weak \xray\ upper limit for J1237+6301. 
The mean \daox\ value for the \hbox{high-redshift}, radio-quiet WLQs is
$-0.144\pm0.075$. As expected the combined \hbox{low-redshift} and
high-redshift, radio-quiet WLQ sample (mean \daox\ value of
$-0.187\pm0.056$) also follows a different \daox\ distribution
from that of typical SDSS quasars (null-hypothesis probability of
$4.5\times10^{-6}$). The \daox\ distribution of
\hbox{low-redshift}, radio-quiet WLQ candidates is consistent with
that of high-redshift, radio-quiet WLQs (null-hypothesis
probability of 0.35), though the sample sizes being compared are
limited.

\subsection{Classifying Radio-Quiet WLQs}\label{discuss:class}

% Figure 7: alpha_ro vs. alpha_ox
\begin{figure*}[t]
    \centering
    \includegraphics[width=6.3in]{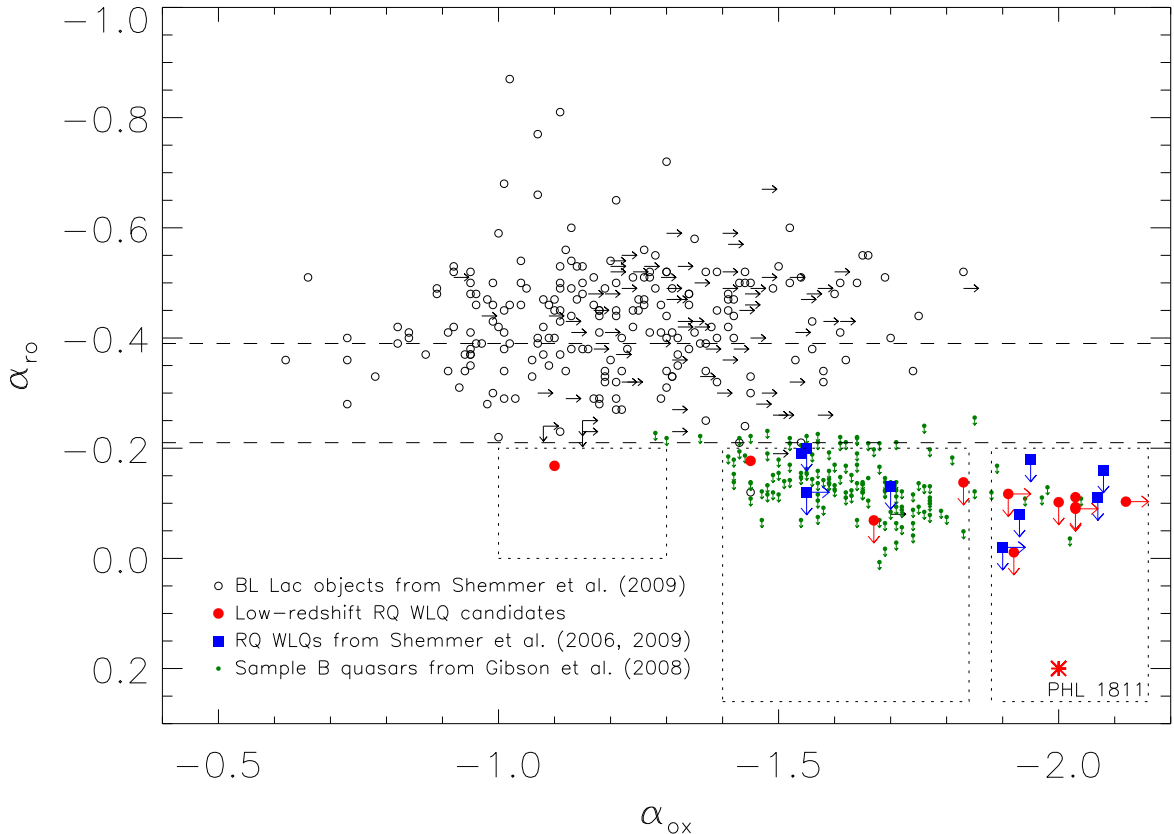}

    \vspace{0.1cm}
    \caption{\footnotesize{$\alpha_{\rm ro}$--$\alpha_{\rm ox}$ diagram for WLQ candidates (red filled circles 
             for \hbox{low-redshift} objects in our sample; blue filled squares for high-redshift objects
             in Shemmer et al. 2006, 2009), BL~Lac objects (black open circles; Shemmer et al. 2009), and 
             typical radio-quiet SDSS quasars (small green dots; Sample~B quasars of Gibson et~al. 2008). 
             The red asterisk represents PHL~1811. Rightward (downward) pointing arrows represent \aox\ 
             (\aro) upper limits. The two 
             dashed lines mark the criteria for radio-quiet (\aro $> -0.21$), radio-intermediate 
             ($-0.39 <$\aro$<-0.21$), and radio-loud (\aro$<-0.39$) objects. The boxes bordered by dotted lines
             show the three suggested groups of WLQ candidates based on their multi-band properties. Note that 
             the WLQ sample has an excess of objects with large negative \aox\ values, compared to both 
             typical radio-quiet quasars and BL~Lac objects.}
             \label{aroaox_fig}}
\end{figure*}

% Figure 8: C IV diagram
\begin{figure*}[t]
    \centering
    \includegraphics[width=6.0in]{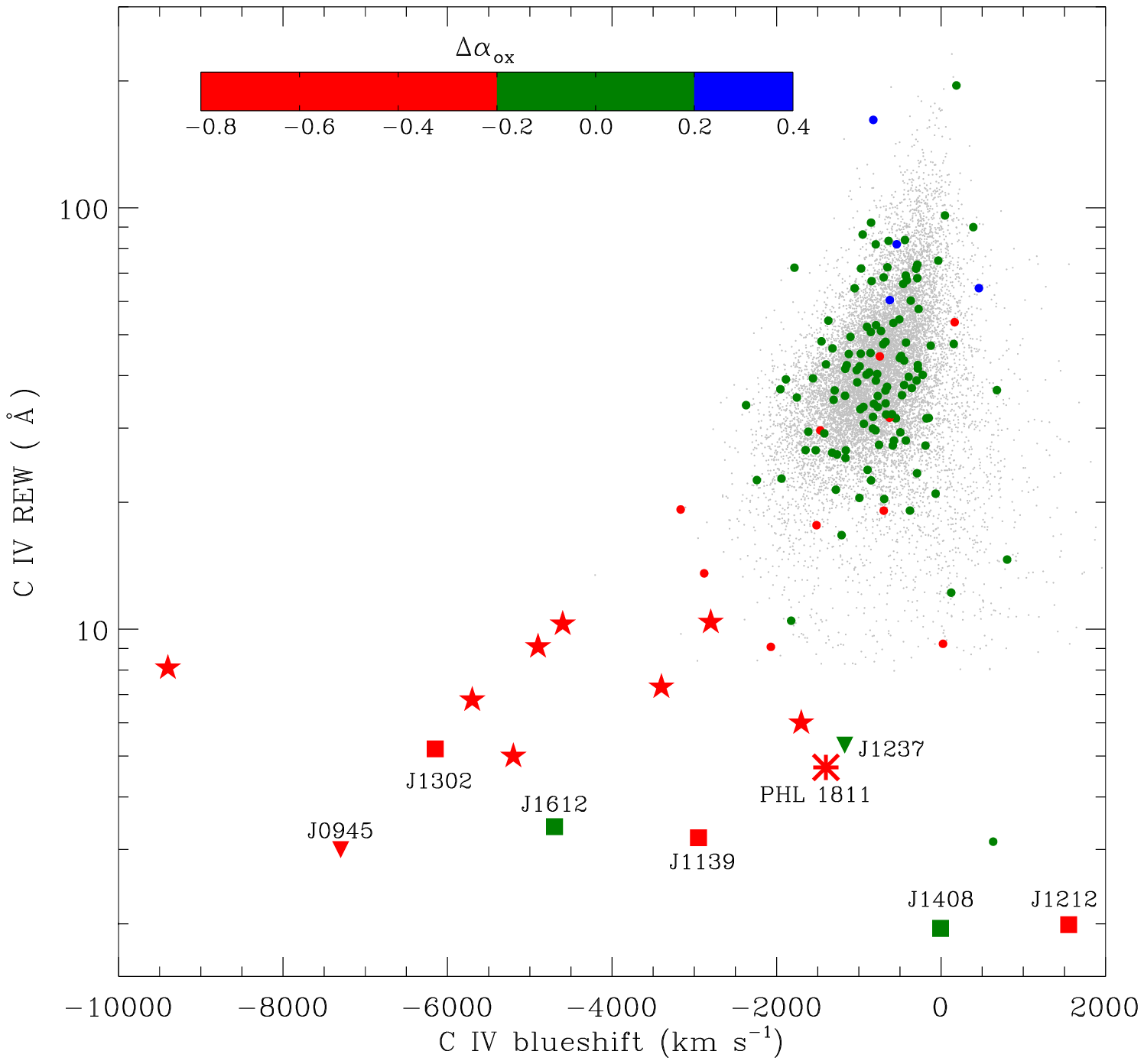}
    \caption{\footnotesize{REW(\civ) plotted against the \civ\ blueshift for our radio-quiet WLQ candidates 
             (filled squares for \xray\ detected sources, filled upside-down triangles for \xray\ undetected 
             sources), \phl\ (asterisk), radio-quiet \phl\ analogs in Wu et~al. (2011) (stars), 
             and \hbox{radio-quiet}, \hbox{non-BAL} quasars in Sample~B of 
             Gibson et~al. (2008a) (circles). These sources are color-coded according to their \daox\ values 
             (three color bins are used, corresponding to the \xray\ weak, \xray\ normal and \xray\ strong 
             sources described in \S\ref{discuss:class}, respectively). The color bar shows the \daox\ 
             range for each color. Source names for 
             WLQs are labeled in the format of 'J$hhmm$' for brevity. The grey dots show the 
             13,582 \hbox{radio-quiet} quasars in Sample A of Richards et~al. (2011; see their Fig.~7).}
             \label{c4bew_fig}}
\end{figure*}

To investigate the multi-band properties of \hbox{low-redshift}, radio-quiet WLQ candidates,
we plotted the sources of our sample in an \aro-\aox\ diagram (Fig.~\ref{aroaox_fig}) along
with the high-redshift, radio-quiet WLQs in Shemmer et~al. (2006, 2009), the BL~Lac sample
in Shemmer et~al. (2009), and the Sample~B quasar in Gibson et~al. (2008). 
The \hbox{low-redshift}, radio-quiet WLQ candidates have similar
multi-band properties to those of high-redshift, radio-quiet WLQs. They are generally much
fainter in radio and \hbox{X-rays} than most of the BL~Lac objects. The weak emission lines
of \hbox{low-redshift}, radio-quiet WLQ candidates are therefore not likely due to the dilution by
relativistically boosted continua as for BL~Lac objects (see discussion in \S4.1 of
Shemmer et~al. 2009). However, it is possible that a small percentage of the WLQ candidates actually belong
to the radio-faint tail of the BL~Lac population (see below).

The population of radio-quiet WLQ candidates (both at low redshift and high redshift) has a wide
dispersion of relative \xray\ brightness and UV emission-line properties. Motivated by their
\daox\ distribution, their emission-line properties discussed below, and observations of related
objects (e.g., Wu et~al. 2011), we will discuss them in three groups.

%appears
%suggestively to form three groups based on their multi-band brightness and emission-line properties (see
%Fig.~\ref{daox_fig} and Fig.~\ref{aroaox_fig}).

The majority of WLQ candidates are not only \xray\ weaker than
BL~Lac objects, but also weaker than typical radio-quiet SDSS
quasars. These WLQ candidates may belong to the notable class of
\xray\ weak quasars termed ``\phl\ analogs'' which were recently studied in
detail by Wu et~al. (2011). The \phl\ analogs generally have weak
and highly blueshifted high-ionization lines (e.g., \civ, Si~{\sc iv}),
weak semi-forbidden lines (e.g., C~{\sc iii}]), and strong UV
Fe~{\sc ii} and/or Fe~{\sc iii} emission. Some of our
\hbox{low-redshift}, radio-quiet WLQs have similar UV
emission-line properties to those of \phl, as listed below:
\begin{enumerate}
\item J0812+5225 (\daox$=-0.42$) has weak C~{\sc iii}] and strong Fe~{\sc ii} emission. 
\item J0945+1009 (\daox$\;<\;-0.34$) has weak \civ\ and C~{\sc iii}] emission lines. Its \civ\ line is
highly blueshifted ($\approx -7000$~km~s$^{-1}$). 
\item J1252+2640 (\daox$=-0.39$) has weak C~{\sc iii}] and strong Fe~{\sc ii} emission.
\item J1139$-$0201 (\daox$=-0.38$) has weak and highly blueshifted ($\approx
-2950$~km~s$^{-1}$) \civ\ emission, weak C~{\sc iii}] emission,
and strong Fe~{\sc iii} emission. 
\end{enumerate}
A high-redshift, radio-quiet
WLQ J1302+0030 (\daox$=-0.38$) also has a weak and highly blueshifted \civ\
emission line (DS09; Wu et~al. 2011). All of the above mentioned
sources are \xray\ weak by a factor of $>\;7$ (see
Table~\ref{aox_table}). Fig.~\ref{c4bew_fig} shows the
distribution of our WLQ candidates in the \daox$-$\ion{C}{4}
blueshift$-$ REW(\ion{C}{4}) parameter space. J0945+1009,
J1139$-$0201 and J1302+0030 are similar to \phl\ analogs in this
diagram. Based on the model in \S4.6 of Wu et~al. (2011), these
\phl\ analogs may have high-ionization shielding gas with large
column density and a large covering factor of the BELR which
blocks most of the ionizing photons, resulting in weak
high-ionization emission lines. If a quasar of this kind is viewed
through the BELR and shielding gas, it would be an \xray\ weak WLQ
with weak and highly blueshifted high-ionization lines (e.g.,
\civ). Based on the estimate in \S4.6 of Wu et~al. (2011), \phl\
analogs should make up $\approx 30\%$ of the total WLQ population.
However, our sample of low-redshift, radio-quiet WLQ candidates
appears to have a higher fraction ($\gtrsim 50\%$) of \phl\ analogs, which may
indicate our sample has some selection bias toward \xray\ weak WLQ
candidates. This bias could perhaps be the result of a more strict
criterion upon the strengths of emission lines for most sources in
our sample (REW~$\lesssim$~5~\AA). Quasars with weaker emission
lines (e.g., \civ) are perhaps more likely to be weak in
\hbox{X-rays} (e.g., see \S4.5 of Wu et~al. 2011). The apparently higher fraction
of \phl\ analogs in our sample than that in Wu et~al. (2011) may perhaps also 
be caused simply by small-sample statistics. A Fisher's exact test (Fisher 1922) 
gives an $11.1\%$ probability for the different fractions of
\phl\ analogs among these two samples under the null hypothesis 
(i.e., the two samples have the same fraction of \phl\ analogs). 

Some of our WLQ candidates have similar \xray\ brightness to that of typical radio-quiet quasars
($-0.2\lesssim$~\daox~$\lesssim0.2$). Their high-ionization lines are also weak, but perhaps
not highly blueshifted (e.g., see Fig.~\ref{c4bew_fig} for J1408+0205). Some of them
(e.g., J1612+5118) have very weak UV Fe~{\sc ii} and/or Fe~{\sc iii} emission. J1612+5118 does 
seem to have a
highly blueshifted \civ\ line, for which the reason is unclear. However, the \civ\ line of
this source is close to the blue border of its SDSS spectral coverage. Further UV spectroscopy
with better \civ\ coverage is needed to confirm its \civ\ blueshift. Based on the model in
Wu et~al. (2011), these sources are similar to \phl\ analogs physically, but they are
viewed at different orientations. These sources are observed along lines of sight that
avoid the shielding gas and the BELR. Therefore they appear normal in \hbox{X-rays}.
Their high-ionization lines are generally not highly blueshifted.

In our WLQ candidate sample, one source (J1109+3736) is remarkably strong in \hbox{X-rays}.
It also shows similar UV/optical spectral properties to those of BL~Lac objects.
This source may belong to the radio-faint tail of the BL~Lac population; we will discuss it
further in \S\ref{discuss:j1109}. It is worth noting that the division of our
radio-quiet WLQ candidates into the three groups discussed above (as shown in
Fig.~\ref{aroaox_fig}) is somewhat arbitrary. We do have some ``border-line'' sources
with \daox$\;\approx\;\pm0.2$ (e.g., J1212+5341). It is difficult to classify these sources
clearly based on current information.

\subsection{The Infrared-to-X-ray SEDs of the Radio-Quiet WLQ Candidates}\label{discuss:seds}

% Figure 9: SEDs
\begin{figure*}[t]
    \centering
    \includegraphics[width=6.2in]{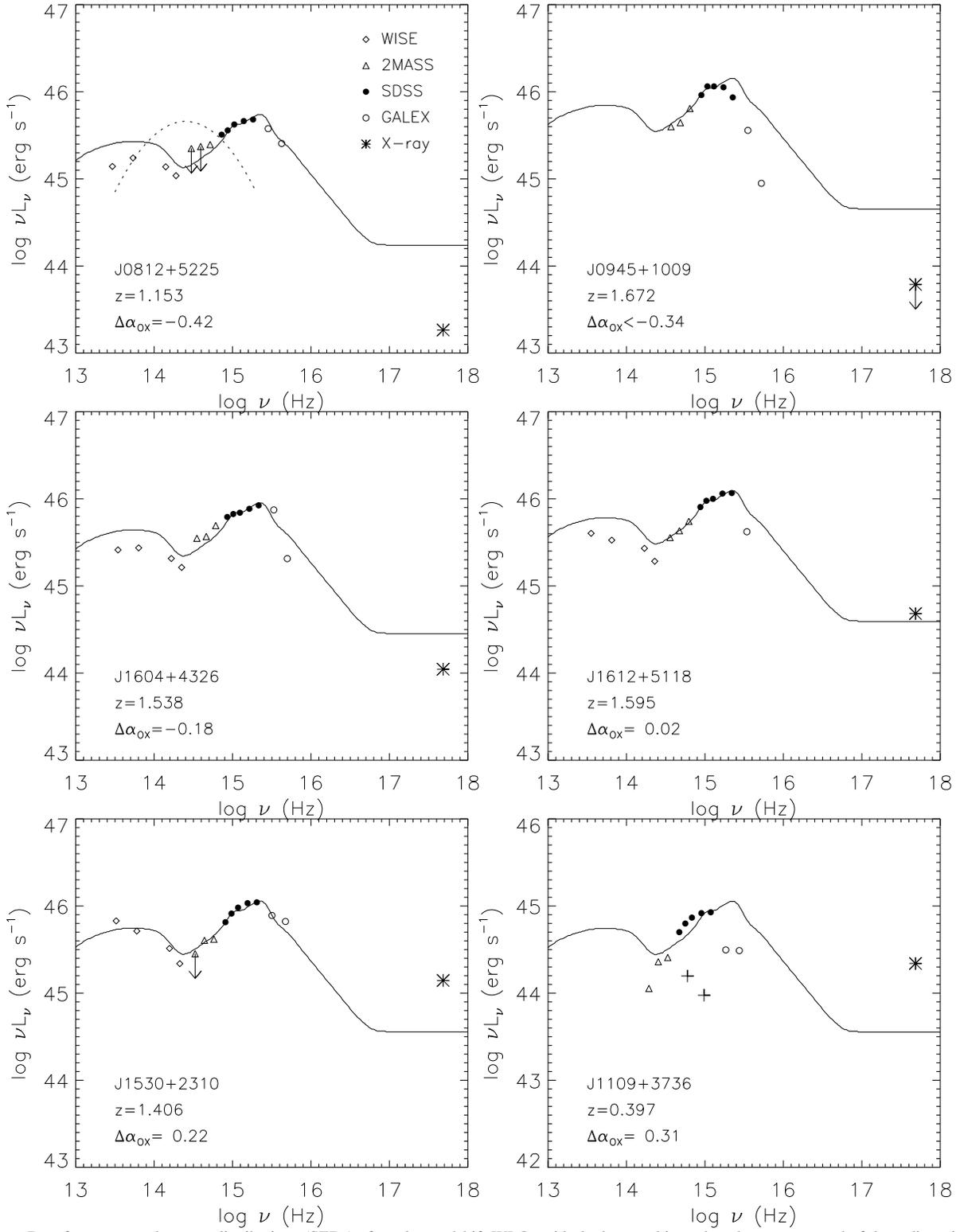}

    \vspace{0.2cm}
    \caption{\footnotesize{Rest-frame spectral energy distributions (SEDs) of our \hbox{low-redshift} WLQs 
             with the best multiwavelength coverage, and of the radio-quiet BL~Lac candidate J1109+3736, 
             ordered by \daox. The photometric data points are from {\it WISE} (open diamonds), 2MASS 
             (open triangles), SDSS (filled circles), {\it GALEX} (open circles), and X-ray observations 
             (asterisks). The average SED of all SDSS quasars from the sample of Richards et al. 
             (2006) is also shown (solid curve), scaled to the flux at rest-frame $10^{15}$~Hz. A parabolic
             SED for typical BL~Lac objects is shown by the dotted line in the top-left panel. The `+' signs
             in the bottom right-panel show the POSS photometry for J1109+3736.}
             \label{sed1_fig}}
\end{figure*}

For the purpose of investigating further the multi-wavelength SEDs of our
\hbox{low-redshift}, radio-quiet WLQs, we gathered photometry for
our sample from the following bands: (1) near- and mid-infrared
from {\it WISE} (The Wide-field Infrared Survey Explorer; Wright
et~al. 2010); (2) near-infrared from 2MASS (The Two Micron All Sky
Survey; Skrutskie et~al. 2006); (3) optical from the SDSS; (4) UV
from {\it GALEX} (The Galaxy Evolution Explorer; Martin et~al.
2005); and (5) \hbox{X-rays} from this work. Fig.~\ref{sed1_fig}
shows the SEDs of the five \hbox{low-redshift} WLQs in our sample
which have the best multi-band coverage. We also include the SED
of the radio-quiet BL~Lac candidate J1109+3736 (see more
discussion in \S\ref{discuss:j1109}). A key point to keep in mind
is that these multi-band observations are non-simultaneous. 
The SEDs are therefore subject to potential distortions due to
variability.

We examined the {\it WISE} and {\it GALEX} image tiles by eye to identify potential 
cases of source blending, confusion, or incorrect matching caused by the low angular 
resolution of {\it WISE} and {\it GALEX}. None of the sources in
our sample is subject to these kinds of problems. The 2MASS magnitudes
in the SDSS DR7 quasar catalog were utilized; this catalog provided aperture photometry for 
additional sources detected down to $2\sigma$ (see \S5 of Schneider et~al. 2010). For the 
sources without detections at $\geq2\sigma$, we adopted flux upper limits obtained 
following the same photometry procedure (C.~M.~Krawczyk \& G.~T.~Richards 2011, private 
communication). The first five sources in Fig.~\ref{sed1_fig}
are from the groups of \xray\ weak and \xray\ normal WLQ candidates
discussed above. Their mid-infrared-to-UV SEDs are generally consistent with the 
composite SEDs of typical SDSS quasars in Richards et~al. (2006), and they are
significantly different from the SEDs of BL~Lac objects (see the dotted parabolic line in the
top-left panel of Fig.~\ref{sed1_fig}; e.g., Nieppola et~al. 2006). We also investigated the 
SEDs for the {\it WISE}-covered radio-quiet objects cataloged in Plotkin et~al. (2010a) which 
do not have sensitive \xray\ coverage (see the Appendix). The majority of them also 
have mid-infrared-to-UV SEDs consistent with those of typical radio-quiet quasars. Lane
et~al. (2011) obtained similar results for their high-redshift WLQs; the composite SED of
their high-redshift WLQs is inconsistent with SEDs of BL~Lac objects.
%\footnote{Since their high-redshift WLQs are \xray\ weaker than BL~Lac objects in general,
%they would be low-energy peaked BL~Lac objects if they indeed belong to the BL~Lac population.}
For one source in our sample, J0945+1009, the flux in the UV band
is lower than for typical SDSS quasars (see the {\it GALEX} data
points in the top-right panel of Fig.~\ref{sed1_fig}). The UV
deficiency of this source may be caused by Ly$\alpha$-forest
intervening absorption. However, Laor \& Davis (2011) argued that
such intervening absorption is not significant ($\sim11\%$ at most)
for this source with $z=1.66$. The near-infrared-to-UV SED of
J0945+1009 can be well fitted with their local black-body model
for a cold accretion disk.

\subsection{X-ray Spectral Properties of \hbox{Low-Redshift} WLQ Candidates}\label{discuss:stack}

Most of the \hbox{low-redshift}, radio-quiet WLQ candidates do not have sufficient \xray\ counts
for an individual \xray\ spectral analysis. We therefore investigate the average \xray\ spectral
properties for these sources via stacking analyses and joint fitting.

A stacked spectral analysis was performed for the six
\hbox{low-redshift}, radio-quiet WLQ candidates with
\daox$\;<\;-0.3$ (J0812+5225, J0945+1009, J1139$-$0201,
J1252+2640, J2115+0001, and J2324+1443). These sources are the
weakest in \hbox{X-rays} among the full sample of
\hbox{low-redshift} WLQ candidates. The detected sources have
similar numbers of \xray\ counts, so that any one of them will not
dominate the stacking analysis. These six sources span a relatively
wide range of redshift ($z=1.15$--$2.50$), and thus the
observed-frame bands of each source correspond to different energy
ranges in the rest frame. However, under the assumption of a simple
power-law spectral model, one can stack the \xray\ counts to
obtain the average effective power-law photon index. We added the \xray\ counts
of these sources in the observed-frame soft band and hard band,
respectively. The numbers of total net counts are
$14.4^{+4.9}_{-3.7}$ in the soft band and $5.2^{+3.4}_{-2.2}$ in
the hard band ($68\%$ confidence level), and the resulting band
ratio is $0.36^{+0.27}_{-0.18}$. With the average Galactic neutral
hydrogen column density of these sources ($N_{\rm
H}=3.50\times10^{20}$~cm$^{-2}$), the band ratio was converted to
an effective power-law photon index $\Gamma=1.66^{+0.63}_{-0.51}$.
The average \xray\ spectrum of the \xray\ weak \hbox{low-redshift}
WLQs is perhaps somewhat harder than that for typical radio-quiet
quasars ($\Gamma\approx2$), but it is consistent within the error
bars. This average \xray\ spectrum is likely softer than that of
the \phl\ analogs at $z=2.19$--$2.38$ ($\Gamma=1.10^{+0.45}_{-0.40}$) in Wu et~al.
(2011) which was also obtained via a stacking analysis, but
consistent within 2$\sigma$. Both stacking analyses suffer from
large uncertainty due to limited \xray\ counts. For a sample combining the six \xray\ 
weak WLQs analyzed here (which are likely to be \phl\ analogs) and the radio-quiet 
\phl\ analogs in Wu et~al. (2011), the average \xray\ spectrum has 
a flat effective power-law photon index of $\Gamma=1.35^{+0.33}_{-0.31}$. Deeper \xray\
observations are necessary to give tighter constraints on the
\xray\ spectral properties of these \xray\ weak WLQ candidates.

Two sources (J0945+1009 and J2115+0001) are undetected by \chandra. Adding the \xray\ counts
of these two sources cannot generate a stacked source that would be detected by \chandra, because
both of these two sources have zero \xray\ counts. However, we are able to obtain a tighter
average constraint on their \xray\ brightness via stacking analysis. The upper limit upon the
soft-band count rate of the stacked source is $2.51\times10^{-4}$~s$^{-1}$. Average values of redshift,
Galactic $N_{\rm H}$, and $f_{2500\mbox{\rm~\scriptsize\AA}}$ are adopted in the following calculation. The
upper limit upon the average flux density at rest-frame 2~keV is
$5.16\times10^{-33}$~erg~cm$^{-2}$~s$^{-1}$~Hz$^{-1}$ under the assumption of the Galactic-absorbed
power law with $\Gamma=2$. The upper limits upon \aox\ and \daox\ are calculated to be \aox~$<\;-2.21$
and \daox~$<\;-0.50$. Therefore, these two sources are \xray\ weak by a factor of $>\;20$ on average.

For the two \xray\ normal, \hbox{low-redshift} WLQs (J1604+4326 and J1612+5118),
we performed joint fitting to
study the average \xray\ spectral properties of these sources. The \xray\ spectra were extracted
from apertures of $3''$ radius centered on the \xray\ positions of these sources via the
standard CIAO routine {\sc psextract}. The background spectra were extracted from annular
regions with inner radii of $6''$ and outer radii of $9''$, which are free of \xray\ sources.
Two spectra were extracted individually for J1604+4326 from its two \chandra\ observations.
Spectral fitting was performed with XSPEC v12.6.0 (Arnaud 1996).
The $C$-statistic (Cash 1979) was used in the spectral fitting instead of
the standard $\chi^2$ statistic because the $C$-statistic is well suited to the limited \xray\
counts in our analysis (e.g., Nousek \& Shue 1989). We fit the spectra jointly using a power-law
model with a Galactic absorption component represented by the \verb+wabs+ model (Morrison
\& McCammon 1983). We also used another model similar to the first, but adding an intrinsic 
(redshifted) neutral absorption component, represented by the \verb+zwabs+ model. Both sources 
were assigned their own values of redshift and Galactic neutral hydrogen column density; 
the Galactic column density was fixed to the values 
calculated with COLDEN (Column~4 of Table~\ref{aox_table}). The joint fitting results
are shown in Table~\ref{xspec_table}.  The quoted errors or upper limits are at the 90\%
confidence level for one parameter of interest ($\Delta\;C=2.71$; Avni 1976; Cash 1979).
The average \xray\ spectral properties of these two sources are similar to those of typical radio-quiet
quasars. They have an average photon index ($\Gamma=2.07^{+0.31}_{-0.30}$), consistent with that
of typical radio-quiet quasars ($\Gamma\approx2$). The average photon
index is also consistent with those from their band-ratio analyses. We did not find evidence
of strong intrinsic neutral absorption for J1604+4326 and J1612+5118
($N_{\rm H}\;\lesssim\;1.58\times10^{22}$~cm$^{-2}$); the spectral fitting quality
was not improved after adding the intrinsic neutral absorption component.

% Figure 11: J1109 spec figures
\begin{figure}[t]
    \centering
    \includegraphics[width=3.0in]{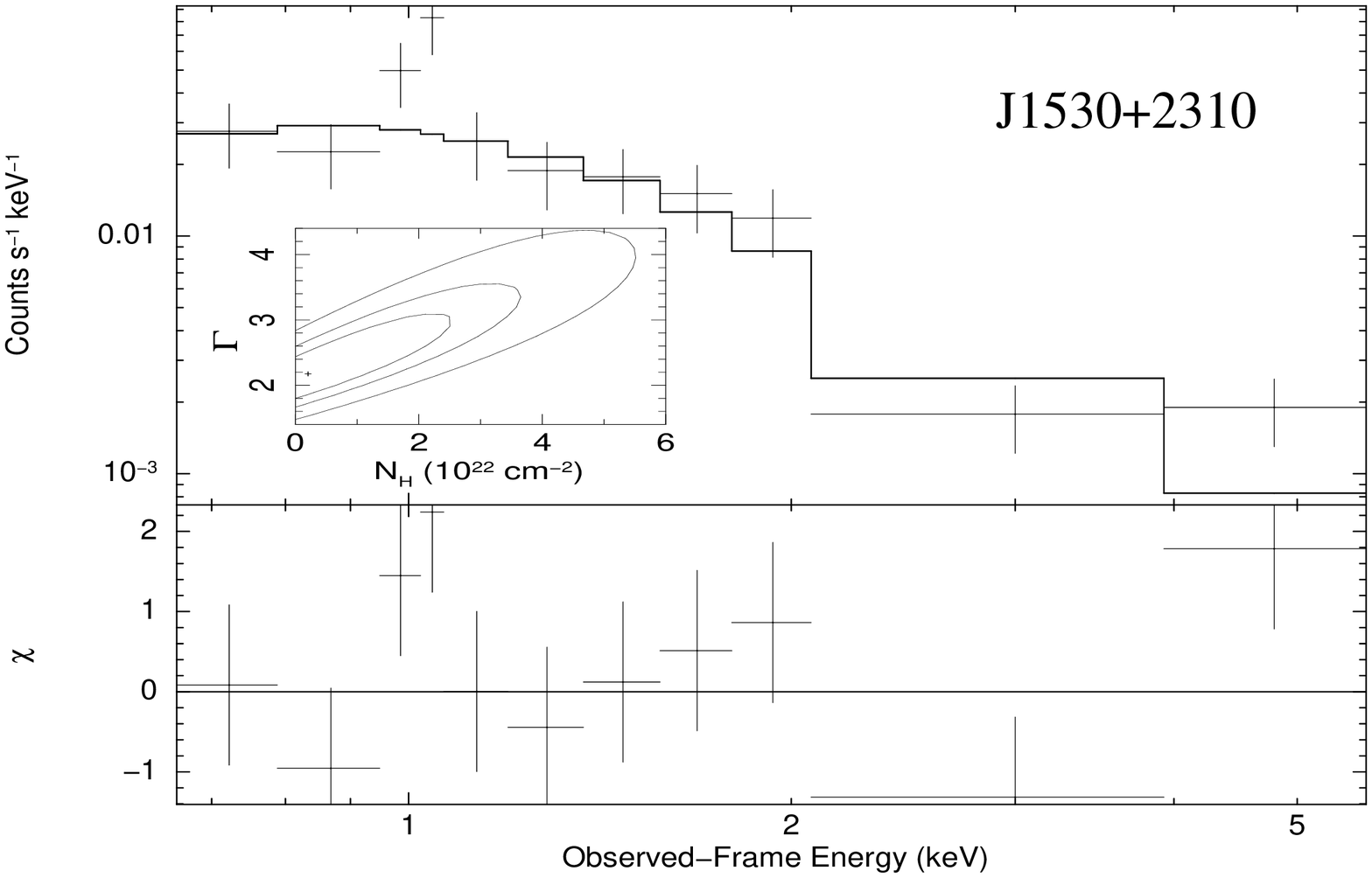}

    \vspace{0.10cm}

    \includegraphics[width=3.0in]{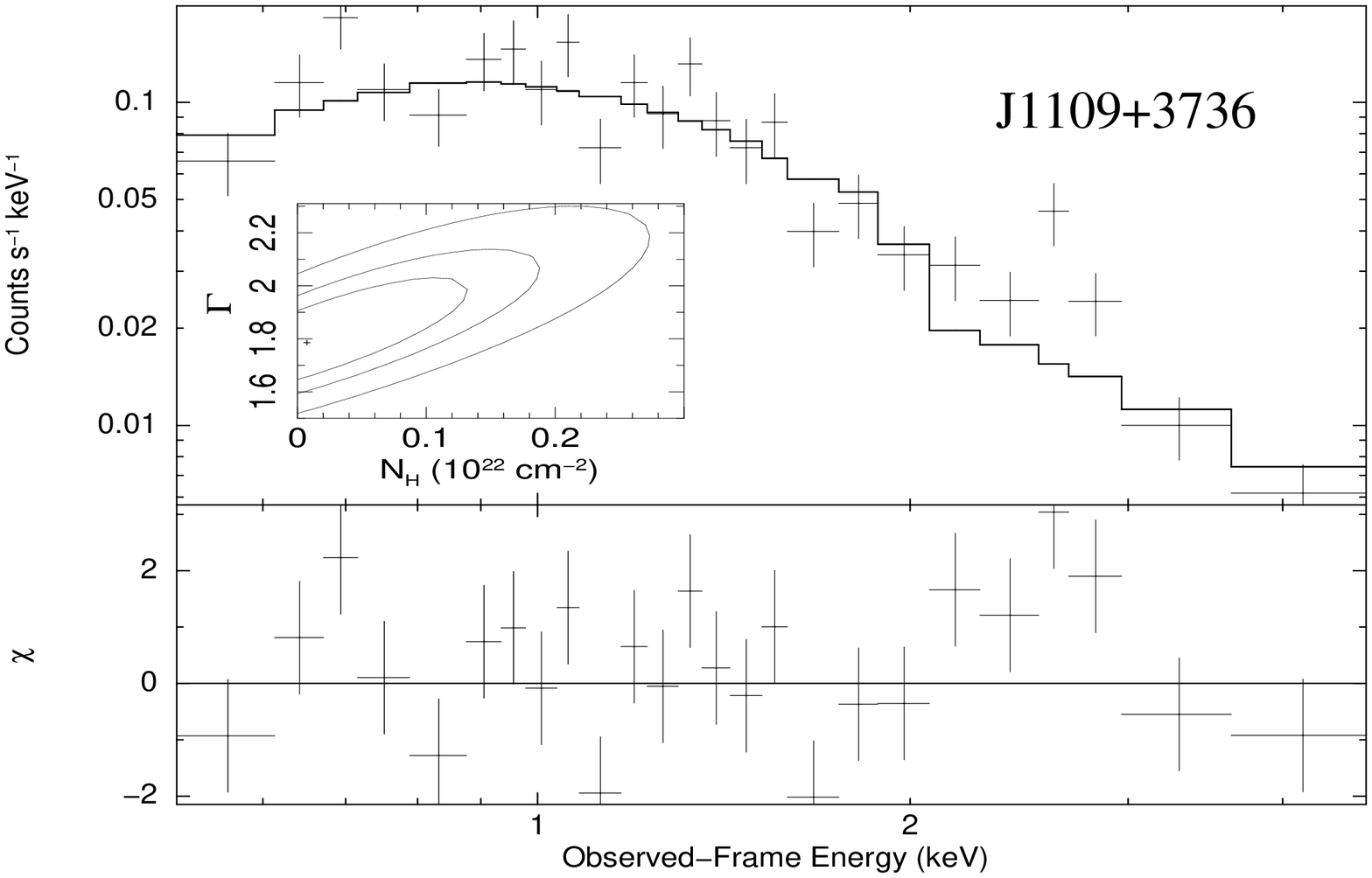}
    \caption{\footnotesize{\hbox{X-ray} spectra of SDSS~J1530+2310 (top panel; binned to a minimum of 
             10 counts per bin) and SDSS~J1109+3736 (bottom panel; binned to a minimum of 20 counts
             per bin) fitted with a power-law model with Galactic absorption. 
             The residuals are shown in units of $\sigma$. The inset of each panel shows contours of
             the photon index vs. intrinsic neutral hydrogen column density parameter space, at confidence
             levels of 68\%, 90\%, and 99\%, respectively.}
             \label{xspec_fig}}
\end{figure}

Two sources (J1109+3736 and J1530+2310) have sufficient \xray\ counts for individual
spectral analysis. We will discuss the \xray\ spectral properties of J1530+2310 here
and leave the radio-quiet BL~Lac candidate J1109+3736 for the next subsection.
The \xray\ spectrum of J1530+2310 was extracted following the same procedure as for
J1604+4326 and J1612+5118 (see above). The spectrum of J1530+2310 was
first grouped to have at least 10 counts per bin (see Fig.~\ref{xspec_fig}). We used the
same spectral models as those for the joint fitting above. The standard $\chi^2$ statistic
was utilized. The fitting results are also shown in Table~\ref{xspec_table}. Fig.~\ref{xspec_fig}
shows the \xray\ spectrum, the best-fit power-law model with Galactic absorption, and the
contour plot of the $\Gamma-N_{\rm H}$ parameter space for the spectral model with intrinsic
neutral absorption for J1530+2310.
The \xray\ spectral properties of J1530+2310 are similar to those of J1604+4326 and J1612+5118.
The photon index of J1530+2310 from the spectral fitting ($\Gamma=2.11^{+0.37}_{-0.34}$) is consistent
with that from the band-ratio analysis (see Table~\ref{cts_table}).
J1530+2310 also shows no evidence of strong intrinsic neutral absorption
($N_{\rm H}\;\lesssim\;2.67\times10^{22}$~cm$^{-2}$). In summary, the \xray\ spectral properties
of \xray\ normal \hbox{low-redshift} WLQs are consistent with those of typical radio-quiet quasars.
Shemmer et~al. (2009) found similar results for their high-redshift, radio-quiet WLQs.
Their sources have a somewhat harder average \xray\ spectrum ($\Gamma=1.86^{+0.72}_{-0.48}$), perhaps
because they fit both \xray\ weak and \xray\ normal WLQs jointly.
Shemmer et~al. (2009, 2010) also performed individual \xray\ spectral analysis on two high-redshift,
radio-intermediate WLQs (J1141+0219 and J1231+0138).

\subsection{J1109+3736: The Radio-quiet BL Lac Candidate}\label{discuss:j1109}

J1109+3736 is a radio-quiet BL~Lac candidate based on its multi-band properties.
It is strong in \hbox{X-rays} by a factor of $\approx6.3$ (\aox$\;=\;-1.10$,
\daox$\;=\;0.31$). Its \aox\ value is
similar to that of the majority of the BL~Lac population (see Fig.~\ref{aroaox_fig}). \xray\
spectral analysis was carried out for J1109+3736 following similar procedures as those in
\S\ref{discuss:stack}. The \chandra\ spectrum of J1109+3736 was grouped to have at least
20 counts per bin (see Fig.~\ref{xspec_fig}). %Standard $\chi^2$ statistic was utilized.
The best-fit parameters are listed in Table~\ref{xspec_table}. The \xray\ spectrum,
the best-fit power-law model with Galactic absorption, and $\Gamma-N_{\rm H}$
contours for J1109+3736 are also shown in Fig.~\ref{xspec_fig}. The best-fit photon index of
J1109+3736 ($\Gamma=1.77\pm0.14$) is consistent with that from the band-ratio analysis. This 
photon index value indicates a harder \xray\ spectrum than those for the majority of the high-energy
peaked BL~Lac objects (HBL),\footnote{J1109+3736 would be classified as
an HBL if it were indeed a BL~Lac object, based on the criterion
$\alpha_{\rm rx}\;<\;0.75$ of Padovani \& Giommi (1995), where $\alpha_{\rm rx}$ is the
radio-to-X-ray spectral index, $\alpha_{\rm rx}=-0.130\;\log(f_{\rm 1~keV}/f_{\rm 5~GHz})$.
The $\alpha_{\rm rx}$ value of J1109+3736 is estimated to be 0.46.} which have a mean
photon index $\Gamma\approx2.2$, but it is still consistent with the broad distribution
of HBL photon-index values (e.g., see the bottom panel of Fig.~1 in Donato et~al. 2005). 
Although this source has a high \xray\ count rate, we do not expect strong photon pile-up 
effects because a $1/2$ subarray mode was used for 
its \chandra\ observation. There is no evidence for significant intrinsic absorption in 
the J1109+3736 spectrum ($N_{\rm H}\;\lesssim\;1.4\times10^{21}$~cm$^{-2}$).

The SDSS spectrum of J1109+3736 (see the bottom-right panel of Fig.~\ref{spec_fig})
shows a strong power-law
continuum without any detectable emission lines, in particular no Balmer lines. Note that
some of the \xray\ weak, \hbox{low-redshift} WLQ candidates (e.g., \phl; see Leighly et~al.
2007b) have fairly strong Balmer lines. The strength of its Ca~{\sc ii} H/K
break\footnote{The strength of the Ca~{\sc ii} H/K break is defined as the fractional change
of the average flux densities in continuum regions blueward (3750--3950~\AA) and redward
(4050--4250~\AA) of the H/K break (see Equation 1 of Plotkin et~al. 2010a).}
($C=0.185$) indicates
a relatively small contribution to the SDSS spectrum from the host galaxy. J1109+3736 shows
moderate \xray\ variability
between its \chandra\ observation and the epoch of the \rosat\ All Sky Survey in 1990
November (RASS; Voges
et~al. 1999). It was not detected by RASS. The upper limit for \aox\ is estimated to
be \aox$\;<\;-1.29$, showing a factor of $>3$ variation compared to its \chandra\
observation. The quasar should have been detected in the RASS if it had the same \xray\ brightness
as that in the \chandra\ epoch (expecting $\approx20$ counts in a 250~s RASS observation). There is also
evidence for optical variability of J1109+3736. This source was optically identified during
the Cambridge APM (Automated Plate Measuring Machine) scans of POSS (Palomar Observatory
Sky Survey) plates (e.g., McMahon et~al.
2002). We converted the APM magnitudes in the $O$ and $E$ bands ($O=20.76,\;E=19.12$) into a
$B$-band magnitude ($B=20.57$) using Equations \hbox{(1--3)} in McMahon et~al. (2002). We
also converted the SDSS magnitudes to a $B$-band magnitude ($B=18.88$) using the equations in
Table~1 of Jester et~al. (2005). The $B$-band magnitude difference is 1.69, indicating a
factor of $\sim 5$ variability over a time span of $\sim50$ years.
This source shows greater variability than most quasars during this time
span (e.g., see Fig.~24 of Sesar et~al. 2006).
%The redshift of J1109+3736 ($z=0.397$) is also similar to the majority of BL~Lac objects.
Therefore, J1109+3736 is a BL~Lac candidate on the radio-faint
tail of the full BL~Lac population. 

The broad-band SED of J1109+3736 is shown in Fig.~\ref{sed1_fig}. It is 
relatively weak in the near infrared compared to its SDSS brightness. However, 
the SED profile of this source could be strongly affected by its variability 
(see the comparison between its SDSS and POSS fluxes in Fig.~\ref{sed1_fig}) 
given its possible BL~Lac nature. Infrared photometry from {\it WISE}, which will be available in 
the {\it WISE} full data release,\footnote{J1109+3736 is
not covered by the currently available {\it WISE} preliminary data
release. See coverage map at
http://wise2.ipac.caltech.edu/docs/release/prelim/.} will be helpful because the
infrared-to-UV SEDs of BL~Lac objects are very different from
those of typical quasars (e.g., Lane et~al. 2011). Future
polarization measurements of J1109+3736 will also be useful in
understanding its nature, since BL~Lac objects are usually highly
polarized in the optical/UV band. However, polarimetry surveys of
optically selected BL~Lac samples (e.g., Smith et~al. 2007; Heidt
\& Nilsson 2011) did not find any highly polarized radio-quiet
BL~Lac candidates, indicating this kind of source, if it indeed
exists, should make up only a small fraction of the total population of radio-quiet WLQ candidates.

% -----------------------------------------------------------------------------
% Summary
% -----------------------------------------------------------------------------

\section{Summary and Future Studies}\label{summary}

We have compiled a sample of 11 radio-quiet WLQ candidates with $z=0.4$--$2.5$ 
and presented their \xray\ and multiwavelength properties. These sources are mainly
selected from the catalog of radio-quiet, weak-featured SDSS quasars in Plotkin
et~al. (2010a). Six of them were observed in new \chandra\ Cycle~12 observations,
while five have archival \chandra\ or \xmm\ coverage. Our main results are
summarized as follows:
\begin{enumerate}
\item All newly observed \hbox{low-redshift}, radio-quiet WLQ candidates are
detected by \chandra, except for J0945+1009. Three sources (J0812+5225, J0945+1009,
and J1252+2640) are \xray\ weak by factors of $\gtrsim\;$8--12 compared to typical 
quasars with similar optical/UV luminosity. Two sources
(J1530+2310 and J1612+5118) are \xray\ normal, while the other one (J1109+3736)
is \xray\ strong by a factor of 6.3. See \S\ref{xray}.
\item Three (J1139$-$0201, J1604+4326, and J2324+1443) of the five \hbox{low-redshift},
radio-quiet WLQ candidates with sensitive archival \xray\ coverage are detected in \hbox{X-rays},
while two (J1013+4927 and J2115+0001) do not have \xray\ detections. All of the five archival
sources are \xray\ weak by factors of $\gtrsim\;$3--12. See \S\ref{xray}.
\item The distribution of relative \xray\ brightness for our \hbox{low-redshift}, radio-quiet WLQ
candidates is significantly different from that of typical radio-quiet quasars. Our sample
has a highly statistically significant excess of \xray\ weak sources. About 64\% (7/11) of the low-redshift, 
radio-quiet WLQs and about 56\% (5/9) of the high-redshift, radio-quiet WLQs are \xray\ weak. The \xray\ weakness
that is commonly found within WLQ samples may well be the driver of the weak broad-line emission. Therefore, 
\xray\ weakness provides an important clue for understanding the nature of WLQs. See \S\ref{discuss:daox}.
\item The \xray\ weak sources (\daox$<-0.2$) in our low-redshift WLQ sample 
are likely to be \phl\ analogs (see Wu et~al. 2011).
Some of them show similar UV emission-line properties to those of \phl\ (weak and highly blueshifted
high-ionization lines, weak semi-forbidden lines, and strong UV Fe emission).
A stacking analysis of these sources indicates
the average effective power-law photon index to be $\Gamma=1.66^{+0.63}_{-0.51}$ (68\% confidence
level). See \S\ref{discuss:class} and \S\ref{discuss:stack}.
%, which is perhaps harder than that for typical radio-quiet quasars.
\item Sources with $-0.2\lesssim$~\daox~$\lesssim0.2$ have multi-band properties that
suggest they are \xray\ normal WLQs. They also
have weak high-ionization lines, while some of them do not have highly blueshifted high-ionization
lines. Some \xray\ normal sources in our sample have weak UV Fe emission.
The average \xray\ spectral properties of these sources
are similar to those of typical SDSS quasars. According to the unification
model in \S4.6 of Wu et~al. (2011), these sources may have a similar geometry and physical nature to the
\phl\ analogs, but with different viewing angles. See \S\ref{discuss:class} and \S\ref{discuss:stack}.
\item The mid-infrared-to-UV SEDs of the \xray\ weak and \xray\ normal \hbox{low-redshift} WLQ candidates
are generally consistent with the composite SEDs of typical SDSS quasars, suggesting that
these sources are not likely to be BL~Lac objects with relativistically boosted continua and diluted
emission lines. See \S\ref{discuss:seds}.
\item The \xray\ strong source J1109+3736 (\daox$>0.3$) may be a radio-quiet BL~Lac object.
It has similar X-ray brightness \hbox{(\aox$=-1.10$)} to those of
typical BL~Lac objects. The SDSS spectrum of J1109+3736 shows
a similar continuum to those of BL~Lac objects. It also has shown
moderate \xray\ (a factor of $>3$) and optical (a factor of
$\approx5$) variability in comparisons with archival data. A \chandra\ spectral analysis using a
Galactic-absorbed power-law model gives $\Gamma=1.77\pm0.14$; thus
its \xray\ spectrum is harder than those of the majority of high-energy
peaked BL~Lac objects. There is no evidence of significant
intrinsic \xray\ absorption for this source. See \S\ref{discuss:j1109}. 
%Further observations, especially the polarization measurements,
%are needed to constrain its physical nature.
\end{enumerate}

Future studies of larger samples of radio-quiet WLQ candidates will be helpful to clarify their
nature. UV spectroscopy covering the \lyanv\ and \civ\ regions of \hbox{low-redshift}, radio-quiet
objects is necessary to study the REW distributions of these two lines, and their REW correlations.
This will provide insights toward a possible universal definition for WLQs at all redshifts, which should
enable systematic studies of larger, unbiased, and more complete samples of WLQ candidates.
For current WLQ studies, it is difficult to perform reliable mid-infrared-to-X-ray SED and correlation 
analyses because of the limited sample size. Further accumulation of high-quality \xray\ data will 
substantially enlarge the sample sizes for such analyses. Deeper \xray\ observations are
required to convert the \xray\ flux upper limits into detections and thus to study the true
overall distribution of
relative \xray\ brightness for radio-quiet WLQ candidates. High-quality \xray\ spectroscopy should be able
to reveal any \xray\ absorbers in \xray\ weak WLQ candidates, clarifying the cause of their
\xray\ weakness and the geometry of these quasars. More accurate measurements of the photon indices
of their hard \xray\ power-law spectra can better constrain the $L/L_{\rm Edd}$ values of WLQ candidates
(e.g., Shemmer et~al. 2008) to see whether extremely high/low $L/L_{\rm Edd}$ is the physical cause
of their weak broad emission lines. Therefore, radio-quiet WLQ candidates will be excellent targets
of future missions with much higher \xray\ spectroscopic capability, e.g., the {\it Advanced Telescope for
High Energy Astrophysics} ({\it ATHENA}).\footnote{http://www.mpe.mpg.de/athena/home.php} 
Near-infrared spectroscopy
covering the H$\beta$ region of the low-redshift WLQ candidates will also allow estimation of
their $L/L_{\rm Edd}$ values (e.g., Shemmer et~al. 2010).

Although most WLQ candidates likely do not have relativistically boosted continua, this work suggests
that a minority of them may belong to the radio-faint tail of the BL~Lac population. Further
growth of the high-quality multiwavelength database, especially in the infrared band, is crucial
to study the broad-band SEDs of WLQ candidates, which could distinguish BL~Lac objects from
WLQs (see \S\ref{discuss:j1109}). Full release of the {\it WISE} source catalog will greatly
benefit SED studies of radio-quiet WLQ candidates. BL~Lac objects often have large-amplitude
variability and high polarization in the UV/optical band. Long-term UV/optical monitoring and
polarimetry of \xray\ strong WLQ candidates will help to identify their nature conclusively.

% -----------------------------------------------------------------------------
% Acknowledgments
% -----------------------------------------------------------------------------

\begin{acknowledgments}

We thank the anonymous referee for providing helpful comments.
We thank C.~M.~Krawczyk and G.~T.~Richards for providing $2MASS$ photometry for SDSS quasars;
R.~R.~Gibson for help with the continuum $S/N$ calculation; and
M.~C.~Eracleous, K.~L.~Luhman, and Y.~Shen for helpful discussions. 
We gratefully acknowledge support from Chandra \xray\ Center grant GO1-12121X 
(J.W., W.N.B.), NASA ADP grant NNX10AC99G (J.W., W.N.B.), and the Southern California 
Center for Galaxy Evolution (A.M.D.).
Funding for the SDSS and SDSS-II has been provided by the Alfred P. Sloan Foundation,
the Participating Institutions, the National Science Foundation, the U.S. Department
of Energy, the National Aeronautics and Space Administration, the Japanese Monbukagakusho,
the Max Planck Society, and the Higher Education Funding Council for England. The SDSS
Web site is http://www.sdss.org/.
The Hobby-Eberly Telescope (HET) is a joint project of
the University of Texas at Austin, the Pennsylvania State University, Stanford University,
Ludwig-Maximilians-UniversitÃ¤t MÃ¼nchen, and Georg-August-Universit\"{a}t G\"{o}ttingen. The HET is
named in honor of its principal benefactors, William P. Hobby and Robert E. Eberly.

\end{acknowledgments}

% -----------------------------------------------------------------------------
% Appendix
% -----------------------------------------------------------------------------

\appendix
\section{Spectral Energy Distributions of Additional Radio-Quiet WLQ Candidates}

The broad-band SEDs of WLQ candidates are able to provide useful insights into their nature. 
Lane et~al. (2011) showed that the mid-infrared-to-UV SEDs of their high-redshift WLQs were 
consistent with those of typical quasars, but were significantly different from those 
of typical BL~Lac objects. In \S\ref{discuss:seds} we have obtained similar results 
for the \xray\ weak and \xray\ normal WLQ candidates in our low-redshift sample 
(see Fig.~\ref{sed1_fig}). However, it is possible to gain insights from 
SED measurements even when sensitive \xray\ data are not available. Therefore, in this 
appendix we will further study the mid-infrared-to-UV 
SEDs of the radio-quiet weak-featured AGNs cataloged in Plotkin et~al. (2010a) that 
do not have sensitive \xray\ coverage. 
%Mid-infrared-to-UV SEDs are able to provide useful insights into the nature of 
%weak-featured AGNs. In \S\ref{discuss:class} we have shown that for the \xray\ weak and
%\xray\ normal WLQ candidates in our sample, their SEDs are consistent with those of 
%typical quasars, but are significantly different from those of typical BL~Lac objects 
%(see Fig.~\ref{sed1_fig}). In this appendix, we will further study the mid-infrared-to-UV 
%SEDs of the radio-quiet weak-featured AGNs cataloged in Plotkin et~al. (2010a) that 
%do not have sensitive \xray\ coverage. 

We cross-correlated the objects in Table~6 of Plotkin et~al. (2010a) to the 
{\it WISE} source catalog in its preliminary data release, obtaining 30 {\it 
WISE}-detected sources (not including the four sources already discussed in 
\S\ref{discuss:seds}). The {\it WISE} coverage of the remaining sources was 
checked with the online image tile look-up 
tool.\footnote{http://irsa.ipac.caltech.edu/wise/applications/WISETiles/wise.html} 
An additional six sources with {\it WISE} coverage were identified. Among these 
six sources, two (J0901+3846 and J1409$-$0000) were 
identified as stars based on their SDSS spectra and proper-motion data; these two 
sources were removed from our SED study. For the remaining four sources, we  
examined their {\it WISE} image tiles. Two of them (J0755+3525 and J1448+2407) have 
{\it WISE} detections in all four bands that lie below the catalog limit. We 
performed aperture photometry
(using a standard $8.25''$ aperture radius) and obtained their fluxes by scaling their
counts in the aperture to those of nearby sources (within $60''$ separation) appearing
in the {\it WISE} catalog. The other two sources (J0857+2342 and J1541+2631) were
not detected in their {\it WISE} images. Following the standard {\it WISE} photometry 
procedure, we calculated their flux upper limits at a 95\% confidence level by adding 
the aperture flux measurement plus two times the uncertainty. For J0857+2342, we could 
only obtain a flux upper limit in the $w3$ band, because there are nearby bright 
sources in its aperture in the other band images. The photometric data 
in the other near-infrared-to-UV wavebands were obtained following the same methods described in 
\S\ref{discuss:seds}. The \xray\ flux limits from the \rosat\ All Sky Survey 
are taken from Table~8 of Plotkin et~al. (2010a). 

The SEDs of the 34 objects are shown in Fig.~\ref{p10ased_fig}. The majority of 
these sources have SEDs consistent with those of typical radio-quiet quasars 
in Richards et~al. (2006), showing they are more likely to be WLQs in nature 
rather than BL~Lac objects. However, there are also several sources with other kinds of SED 
profiles, which we discuss in more detail below. 

{\it J0834+5112, J1556+3854, J1610+3039, and J1658+6118} --- These four sources 
have very red SEDs (strong in the infrared and weak in the optical/UV). They also have 
very red SDSS spectra. They are more likely to be absorbed quasars than bona fide WLQs. 

{\it J1421+0522} --- This source has an SED profile peaking in the near-infrared band, which is
more similar to the SEDs of BL~Lac objects. 

{\it J1522+4137} --- The SED of this source appears similar to that of J0945+1009, 
which can be fit with a cold accretion disk model as in Laor \& Davis (2009). It is 
probably a high-redshift, radio-quiet WLQ. 

{\it J1633+4227} --- The SED of this source peaks in the near-infrared band and drops
rapidly in both blueward and redward directions. This SED profile is similar 
to that of a radio-loud BL~Lac object with significant host-galaxy 
contamination (e.g., J0823+1524 in Plotkin et~al. 2011). J1633+4227 also has the same 
Ca~{\sc ii} H/K break value ($C=0.33$) as that of J0823+1524. It is possible 
that J1633+4227 is a radio-quiet BL~Lac object with substantial host-galaxy 
contamination, although its radio-loudness needs to be constrained more tightly 
(currently \aro$>\;-0.19$ estimated from its {\it FIRST} coverage). The 
near-infrared peak of the SED of J1633+4227 could also perhaps be caused by extreme variability.

% -----------------------------------------------------------------------------
% Bibliography
% -----------------------------------------------------------------------------

% Figure A1_1: SEDs
\begin{figure*}[t]
    \figurenum{A1}
    \centering
    \includegraphics[width=6.3in]{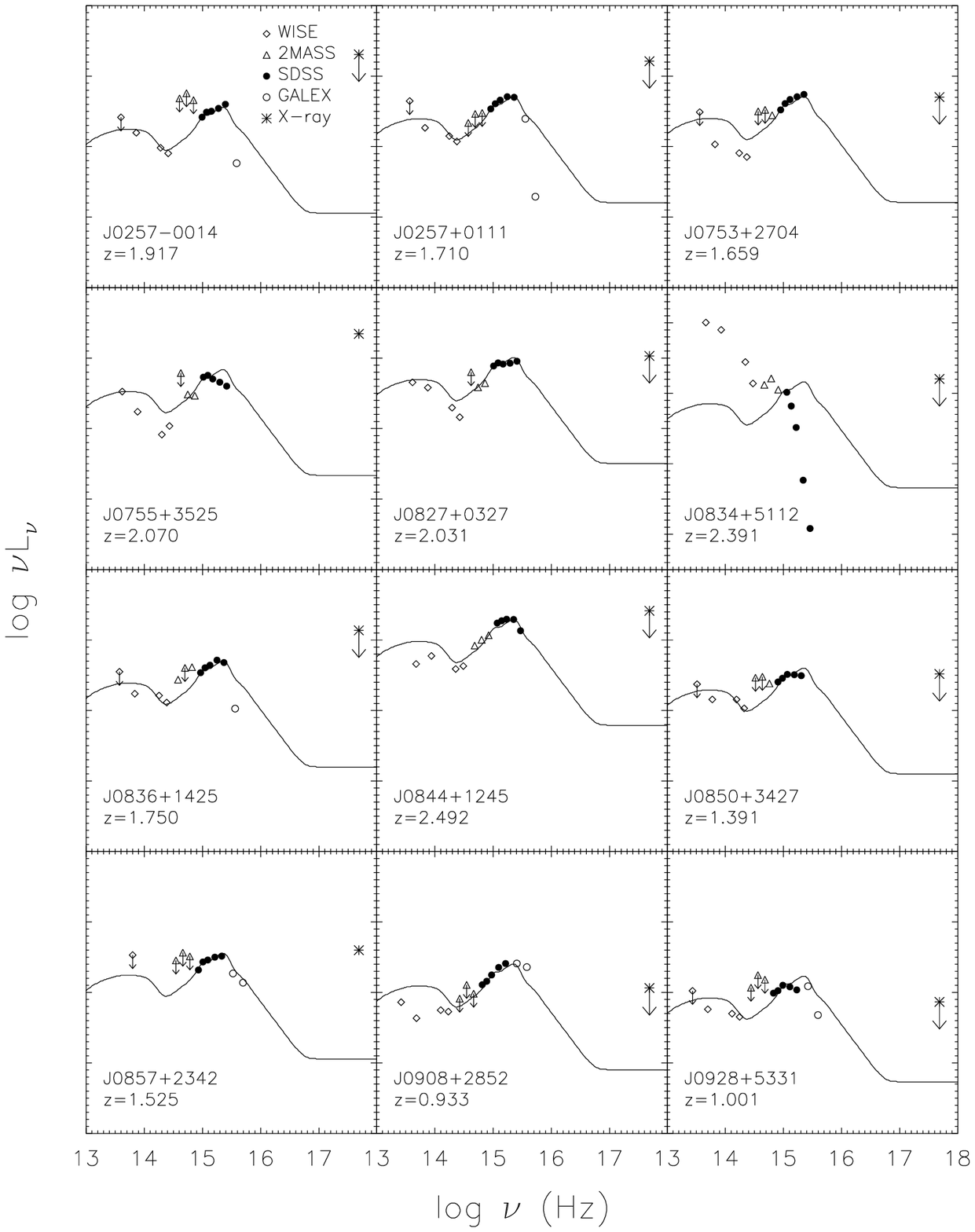}
    \caption{\footnotesize{Spectral energy distributions (SEDs) of additional radio-quiet WLQ 
             candidates in Plotkin et~al. (2010a) with {\it WISE} coverage. All the lines and 
             symbols follow the same definitions as in Fig.~\ref{sed1_fig}. The $y$-axis is in 
             arbitrary units. All the SEDs are in the rest frame, except those for the two 
             sources without redshift information (J1541+2631 and J1556+3854) which are plotted 
             in the observed frame. }
             \label{p10ased_fig}}
\end{figure*}

% Figure A1_2: SEDs
\begin{figure*}[t]
    \figurenum{A1}
    \centering
    \includegraphics[width=6.3in]{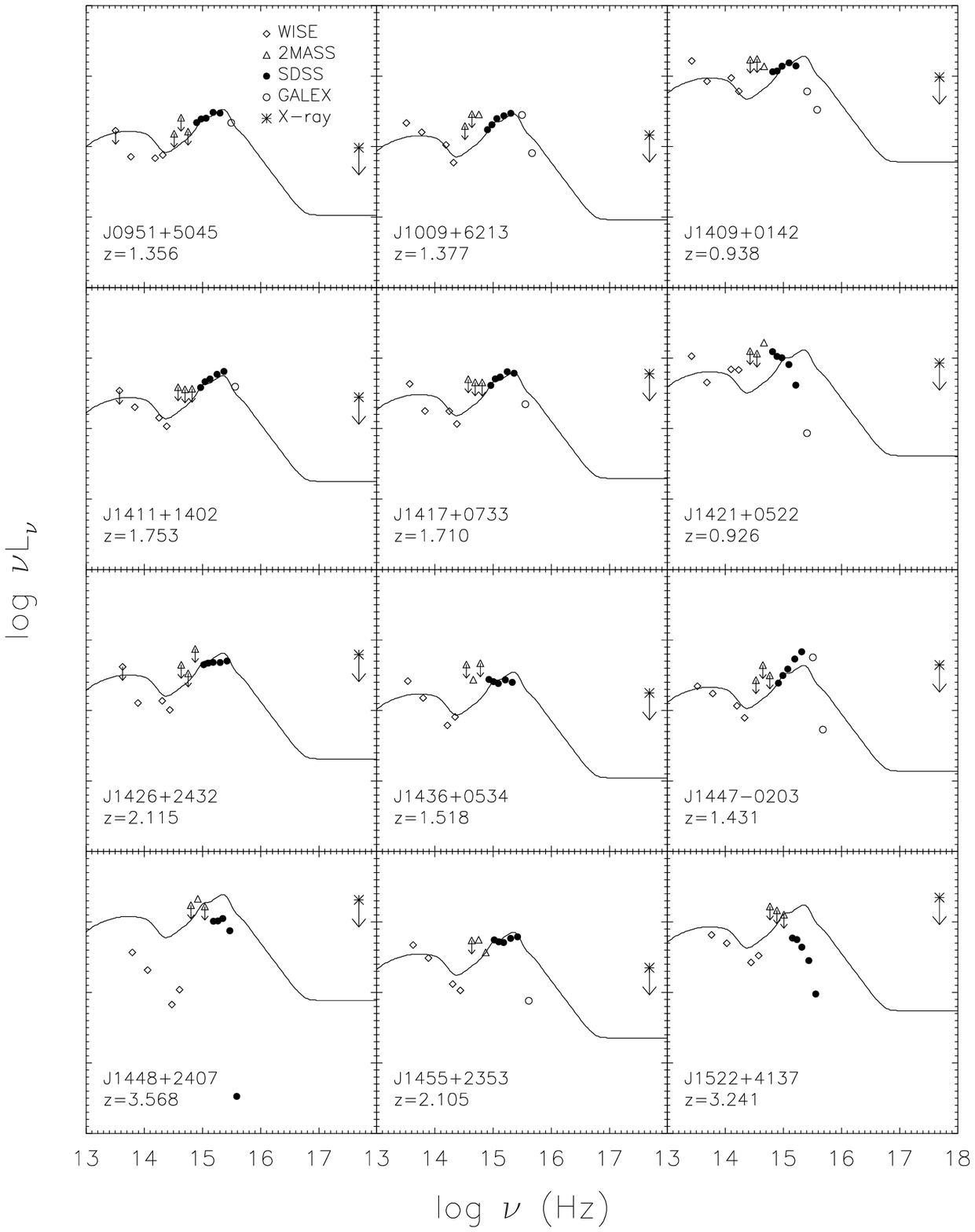}
    \caption{\footnotesize{\it Continued.}
             \label{p10ased2_fig}}
\end{figure*}

% Figure A1_3: SEDs
\begin{figure*}[t]
    \figurenum{A1}
    \centering
    \includegraphics[width=6.3in]{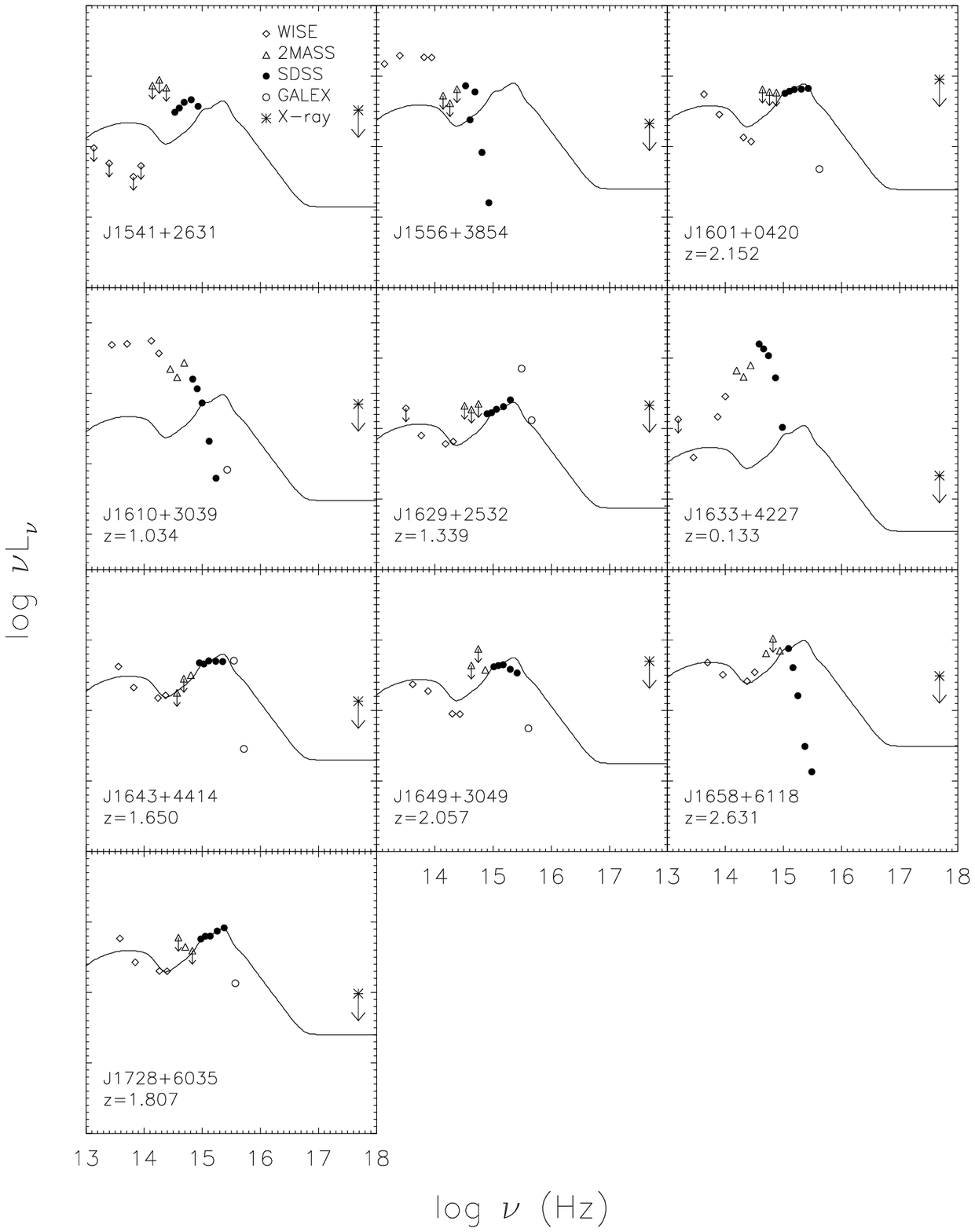}
    \caption{\footnotesize{\it Continued.}
             \label{p10ased3_fig}}
\end{figure*}

% -----------------------------------------------------------------------------
% Tables
% -----------------------------------------------------------------------------

\newpage
% Table 1: Observing Log
\begin{center}
\begin{deluxetable}{clccllcccc}
\tablecolumns{11} \tabletypesize{\scriptsize}
\tablewidth{0pt}
\tablecaption{X-ray Observation Log\label{log_table}}
\tablehead{
    \colhead{} & \colhead{} & \colhead{} & \colhead{$\Delta_{{\rm Opt-X}}$\tablenotemark{b}} & \colhead{Detector} 
               & \colhead{Observation} & \colhead{Observation} & \colhead{Exp. Time} & \colhead{Off-axis Angle} 
               & \colhead{} \\
    \colhead{} & \colhead{Object Name (SDSS~J)} & \colhead{$z$\tablenotemark{a}}  & \colhead{(arcsec)} &
               & \colhead{Date}  & \colhead{ID} & \colhead{(ks)} & \colhead{(arcmin)} & \colhead{References}
}
\startdata
\multicolumn{2}{l}{{\it Chandra} Cycle 12 Objects} & & & & & &\\
& $081250.79+522530.8$ & $1.153$ & $0.8$ & ACIS-S & 2010 Dec 28 & 12710 & $4.1$ & $0.3$  & 1\\	
& $094533.98+100950.1$ & $1.671$ & \nodata & ACIS-S & 2011 Jan 12 & 12706 & $3.0$ & $0.3$  & 2 \\
& $110938.50+373611.7$ & $0.397$ & $0.3$ & ACIS-S & 2011 Feb 27 & 12711 & $3.1$ & $0.3$  & 1\\
& $125219.47+264053.9$ & $1.289$ & $0.3$ & ACIS-S & 2011 Mar 12 & 12709 & $3.4$ & $0.3$  & 1\\
& $153044.08+231013.4$ & $1.406$ & $0.4$ & ACIS-S & 2011 Apr 15 & 12707 & $3.0$ & $0.3$  & 1\\
& $161245.68+511816.9$ & $1.595$ & $0.4$ & ACIS-S & 2011 Feb 01 & 12708 & $3.2$ & $0.3$  & 1\\
\\
\multicolumn{2}{l}{Archival \xray\ Data Objects} & & & & & & \\
& $101353.46+492758.1$ & $1.640$ & \nodata & MOS\tablenotemark{c} & 2004 Apr 23 & 0206340201 & $22.7$ & $6.4$ & 1 \\
& $113900.55-020140.0$ & $1.903$ & $0.2$ & ACIS-S & 2004 Jul 21 & 4871 & $14.9$ & $0.6$ & 1 \\
& $160410.22+432614.6$ & $1.538$ & $0.2$ & ACIS-I & 2006 Jun 25 & 6933 & $26.7$ & $3.7$ & 1 \\
& & & $0.2$ & ACIS-I & 2006 Jun 23 & 7343 & $19.4$ & $3.7$ & \\
& $211552.88+000115.5$ & $2.500$ & \nodata & ACIS-S & 2008 Dec 24 & 10388 & $9.5$ & $0.3$ & 1,3,4\\
%& $224749.57+134248.1$ & $1.175$ & $0.2$ & ACIS-S & 2009 Aug 06 & 10387 & $3.3$ & $0.3$ & 3,4\\
& $232428.43+144324.3$ & $1.417$ & $0.7$ & ACIS-S & 2009 May 31 & 10386 & $5.0$ & $0.3$ & 1,3,4
\enddata
\tablenotetext{a}{Redshift for each source. See \S\ref{uvo:line} for details about redshift measurements.}
\tablenotetext{b}{Angular distance between the optical and \xray\ positions; no entry indicates no \xray\ detection.}
%\tablenotetext{c}{The exposure times are corrected for detector dead time.}
\tablenotetext{c}{This object was observed by both the MOS and pn detectors. We list MOS detector parameters here.}
\tablerefs{(1) Plotkin et~al. 2010a; (2) Hryniewicz et~al. 2010; (3) Collinge et~al. 2005; (4) Plotkin et al. 2010b.}
\end{deluxetable}
\end{center}

%\clearpage
% Table 2: UV emission line measurements
%\begin{turnpage}
\begin{deluxetable}{ccccccccc}
\tabletypesize{\scriptsize}
\tablecaption{Quasar UV Emission-Line Measurements\label{qso_table}}
\tablewidth{0pt}
\tablehead{ \colhead{} &
\colhead{Object Name} & \colhead{MJD} & 
\colhead{C\,{\sc iv} Blueshift} 
& \colhead{REW} & \colhead{REW} & \colhead{REW} & \colhead{REW} & \colhead{REW}\\
\colhead{} &
\colhead{(SDSS~J)} & \colhead{} & 
\colhead{} 
& \colhead{(C\,{\sc iv})} & \colhead{(Si\,{\sc iv})\tablenotemark{a}} & \colhead{($\lambda1900$~\AA)\tablenotemark{b}} & \colhead{(Fe\,{\sc iii})} & \colhead{(Mg\,{\sc ii})}%& \colhead{$\alpha_\nu$\tablenotemark{c}}
}
\startdata
\multicolumn{2}{l}{{\it Chandra} Cycle 12 Objects} & & & & &\\
& $081250.79+522530.8$ & 53297 & \nodata & \nodata & \nodata & $ 3.8\pm 2.1 $ & $< 4.5$ & $8.4\pm 0.7$ \\
& $094533.98+100950.1$ & 52757 & $-7300\pm 1700$ & $3.0\pm 1.2$ & \nodata & $ 4.9\pm 1.5 $ & $1.7\pm 1.5$ & $17.1\pm 0.8$ \\
& $110938.50+373611.7$ & 53499 & \nodata & \nodata & \nodata & \nodata & \nodata & \nodata \\
& $125219.47+264053.9$ & 53823 & \nodata & \nodata & \nodata & $ 8.8\pm 1.2 $ & $ 2.9\pm 0.9$ & $8.7\pm 0.4$ \\
& $153044.08+231013.4$ & 53878 & \nodata & \nodata & \nodata & $ 4.9\pm 1.5 $ & $ 2.6\pm 1.2$ & $12.95\pm 0.4$ \\
& $161245.68+511816.9$ & 52051 & $-4700\pm 1300$ & $3.4\pm 1.8$ & \nodata & $5.1\pm 1.5$ & $ 3.0\pm 1.5 $ & $9.5\pm 0.6$ \\
\\
\multicolumn{2}{l}{Archival \xray\ Data Objects} & & & & & \\
& $101353.46+492758.1$ & 52076 & \nodata & \nodata & \nodata & $ 3.4\pm 1.8 $ & $< 6.9$ & $6.2\pm 0.8$ \\
& $113900.55-020140.0$ & 52294 & \nodata & $< 9.0$ & $< 9.9$ & $< 10.8$ & $< 9.0$ & $11.1\pm 1.1$ \\
& $113900.55-020140.0$ (HET) & 55702 & $-2950\pm 1550$ & $3.2\pm 2.7$ & \nodata & $11.7\pm1.8$ & $5.5\pm1.5$ & \nodata \\
& $160410.22+432614.6$ & 52756 & \nodata & \nodata & \nodata & $< 1.9$ & $< 1.8$ & $5.8\pm 1.0$ \\
& $211552.88+000115.5$ & 52443 & \nodata & \nodata & \nodata & \nodata & \nodata & \nodata \\
& $232428.43+144324.3$ & 52258 & \nodata & \nodata & $7.6\pm 2.1$ & $< 5.4$ & \nodata & $8.6\pm 0.8$ \\
\\
& PHL~1811\ \tablenotemark{c} & \nodata & $ -1400 \pm  250 $ & $  4.7 \pm 0.9 $ & $ 4.8 \pm 0.9 $ & $  8.3 \pm 0.6 $ & $ 4.7 \pm 0.6 $ & \nodata \\
& V01~composite\ \tablenotemark{c,{\rm d}} & \nodata & $  -570 \pm  30 $ & $ 30.0 \pm 0.3 $ &	$ 8.7 \pm 0.3 $ &	$ 21.7 \pm 0.2 $ &	$ 2.9 \pm 0.1 $ & \nodata \\
%Y04~composite~B4\ \tablenotemark{g} & \nodata & $ -470 \pm  260 $ &	$  4380 \pm 250 $ &	$ 4250 \pm  240 $ & $ 1.03 \pm 0.08 $ &	$ 26.0 \pm 0.3 $ &	$ 8.2 \pm 0.3 $ &	$ 22.4 \pm 0.2 $ &	$ 3.1 \pm 0.1 $ & $-0.42$ \\
\enddata
\tablecomments{The blueshift values are in units of km s$^{-1}$. All REW values are in units of \AA.}
%\tablenotetext{a}{The ratio of C\,{\sc iv} FWHM to C\,{\sc iv} $\sigma_{\rm line}$.}
\tablenotetext{a}{This line is a blend of Si~{\sc iv} and O~{\sc iv}]; we refer to it as Si~{\sc iv}
simply for convenience.}
\tablenotetext{b}{Mainly C\,{\sc iii}] $\lambda 1909$, but also including [Ne\,{\sc iii}] $\lambda 1814$, 
Si\,{\sc ii} $\lambda 1816$, Al\,{\sc iii} $\lambda 1857$, Si\,{\sc iii}] $\lambda 1892$, 
and several Fe\,{\sc iii} multiplets (see Table 2 of Vanden Berk et al. 2001).}
%\tablenotetext{c}{The spectral index of the presumed \pl\ continuum, where \hbox{$f_\nu \propto \nu^{\alpha_\nu}$}.}
%\tablenotetext{d}{All the spectral parameters for J1454+0324, including MJD, are the average of two spectra taken at MJD=52029 and MJD=52045.}
%\tablenotetext{e}{The average values for the \phl\ analogs plus \phl\ itself. }
\tablenotetext{c}{These measurements are taken from Wu et~al. (2011).}
\tablenotetext{d}{The composite spectrum from Vanden Berk et al. (2001).}
%\tablenotetext{g}{The `B4' composite spectrum from Yip et al. (2004).}
%\tablenotetext{h}{We measured the emission-line properties for PHL~1092 using its {\it Hubble Space Telescope} STIS spectrum (Leighly et~al. 2007b). See Footnote~13 for further discussion.}
\end{deluxetable}
%\end{turnpage}

\clearpage
% Table 3: Continuum S/N
\begin{turnpage}
%\clearpage
\begin{deluxetable}{ccccccccccccccccccccc}
\tablecolumns{21} \tabletypesize{\scriptsize}
\tablewidth{0pt}
\tablecaption{Continuum $S/N$ Values for SDSS DR7 Quasars\label{contsn_table}}
\tablehead{
    \colhead{} 
    & \colhead{}
    & \colhead{}
    & \colhead{}
    & \colhead{}
    & \colhead{}
    & \colhead{}
    & \multicolumn{4}{c}{$SN_{1700}$ (1650--1750 \AA)\tablenotemark{c}}
    & \colhead{}
    & \multicolumn{4}{c}{$SN_{3000}$ (2950--3050 \AA)\tablenotemark{c}}
    & \colhead{}
    & \multicolumn{4}{c}{$SN_{5150}$ (5100--5200 \AA)\tablenotemark{c}}\\
    \cline{8-11}\cline{13-16}\cline{18-21}
    \colhead{Object Name (SDSS J)} 
    & \colhead{$m_i$\tablenotemark{a}}
    & \colhead{$z$\tablenotemark{b}}
    & \colhead{MJD}
    & \colhead{Plate}
    & \colhead{Fiber}
    & \colhead{}
    & \colhead{$S/N$}
    & \colhead{$N_{\rm pix}$\tablenotemark{d}}
    & \colhead{$\lambda_{\rm lo, obs}$\tablenotemark{e}}
    & \colhead{$\lambda_{\rm hi, obs}$\tablenotemark{f}}
    & \colhead{}
    & \colhead{$S/N$}
    & \colhead{$N_{\rm pix}$\tablenotemark{d}}
    & \colhead{$\lambda_{\rm lo, obs}$\tablenotemark{e}}
    & \colhead{$\lambda_{\rm hi, obs}$\tablenotemark{f}}
    & \colhead{}
    & \colhead{$S/N$}
    & \colhead{$N_{\rm pix}$\tablenotemark{d}}
    & \colhead{$\lambda_{\rm lo, obs}$\tablenotemark{e}}
    & \colhead{$\lambda_{\rm hi, obs}$\tablenotemark{f}}
}
\startdata
$000006.53+003055.2$ &  $20.09$ & $1.825$ & $52203$ & $685$ & $467$ & &   $4.20$ & $257$ & $4660$ & $4943$ & &  $2.22$ & $146$ & $8333$ & $8616$ & &   $0.00$ &   $0$ &    $0$  &  $0$ \\
$000008.13+001634.6$ &  $19.49$ & $1.837$ & $52203$ & $685$ & $470$ & &   $4.83$ & $256$ & $4682$ & $4965$ & &  $3.87$ & $146$ & $8370$ & $8654$ & &   $0.00$ &   $0$ &    $0$  &  $0$ \\
$000009.26+151754.5$ &  $19.15$ & $1.199$ & $52251$ & $751$ & $354$ & &   $2.30$ & $ 57$ & $3798$ & $3848$ & & 1$0.97$ & $146$ & $6485$ & $6705$ & &   $0.00$ &   $0$ &    $0$  &  $0$ \\
$000009.38+135618.4$ &  $18.30$ & $2.234$ & $52235$ & $750$ & $ 82$ & &  1$4.22$ & $256$ & $5337$ & $5660$ & &  $0.00$ & $  0$ & $   0$ & $   0$ & &   $0.00$ &   $0$ &    $0$  &  $0$ \\
$000009.42-102751.9$ &  $18.77$ & $1.845$ & $52143$ & $650$ & $199$ & &   $9.64$ & $256$ & $4695$ & $4979$ & &  $5.61$ & $146$ & $8393$ & $8678$ & &   $0.00$ &   $0$ &    $0$  &  $0$ 
\enddata
\tablecomments{This table is available in its entirety in the online journal. Part of the table is shown here for guidance about its format.}
\tablenotetext{a}{ The apparent $i$-band magnitude using the BEST photometry of the SDSS DR7 quasar catalog.}
\tablenotetext{b}{ The redshift value from the SDSS DR7 quasar catalog.}
\tablenotetext{c}{ Wavelength range in the rest frame.}
\tablenotetext{d}{ Number of pixels used in the calculation.}
\tablenotetext{e}{ The lower limit of the observed-frame wavelength range used in the calculation.}
\tablenotetext{f}{ The upper limit of the observed-frame wavelength range used in the calculation.}
\end{deluxetable}
\end{turnpage}

\clearpage
% Table 4: X-ray photometry
\begin{center}
%\clearpage
\begin{deluxetable}{llccccc}
\tablecolumns{7} \tabletypesize{\footnotesize}
%\tablenum{4}
\tablewidth{0pt}
\tablecaption{X-ray Counts and Basic Spectral Properties\label{cts_table}}
\tablehead{
    \colhead{}     
    & \colhead{}                     
    & \colhead{Full Band}     
    & \colhead{Soft Band}     
    & \colhead{Hard Band}     
    & \colhead{Band}
    & \colhead{}\\
    \colhead{}
    & \colhead{Object Name (SDSS J)} 
    & \colhead{(0.5--8.0 keV)\tablenotemark{a}}
    & \colhead{(0.5--2.0 keV)\tablenotemark{a}} 
    & \colhead{(2.0--8.0 keV)\tablenotemark{a}} 
    & \colhead{Ratio\tablenotemark{b}}  
    & \colhead{$\Gamma$\tablenotemark{c}}
}
\startdata
\multicolumn{2}{l}{{\it Chandra} Cycle 12 Objects} & & & \\
& $081250.79+522530.8$ & $3.3^{+3.0}_{-1.7}$ & $3.1^{+3.0}_{-1.7}$ & $<3.3$ & $<1.06$ & $>0.72$ \\
& $094533.98+100950.1$ & $<3.3$ & $<3.2$ & $<3.3$ & \nodata & \nodata \\
& $110938.50+373611.7$ & $552.3^{+24.5}_{-23.5}$ & $406.4^{+21.2}_{-20.2}$ & $145.8^{+13.1}_{-12.1}$ & $0.36^{+0.04}_{-0.03} $ & $1.68^{+0.09}_{-0.09} $ \\
& $125219.47+264053.9$ & $5.5^{+3.5}_{-2.3}$ & $3.1^{+3.0}_{-1.7}$ & $<7.1$ & $<2.25$ & $>-0.01$ \\
& $153044.08+231013.4$ & $125.8^{+12.2}_{-11.2}$ & $101.9^{+11.1}_{-10.1}$ & $23.9^{+6.0}_{-4.9}$ & $0.23^{+0.06}_{-0.05} $ & $2.12^{+0.22}_{-0.21} $ \\
& $161245.68+511816.9$ & $39.3^{+7.3}_{-6.2}$ & $30.5^{+6.6}_{-5.5}$ & $7.2^{+3.8}_{-2.6}$ & $0.23^{+0.14}_{-0.10} $ & $2.05^{+0.45}_{-0.40} $ \\
\\
\multicolumn{2}{l}{Archival \xray\ Data Objects} & & & \\
& $101353.46+492758.1$ & $<38.6$ & $<25.7$ & $<28.8$ & \nodata & \nodata \\
& $113900.55-020140.0$ & $5.3^{+3.5}_{-2.2}$ & $4.1^{+3.2}_{-1.9}$ & $<5.1$ & $<1.25$ & $>0.55$ \\
& $160410.22+432614.6$ & $46.5^{+7.9}_{-6.8}$ & $41.3^{+7.5}_{-6.4}$ & $8.2^{+4.0}_{-2.8}$ & $0.20^{+0.10}_{-0.07} $ & $2.21^{+0.42}_{-0.38} $ \\
& & $31.5^{+6.7}_{-5.6}$ & $26.3^{+6.2}_{-5.1}$ & $<14.8$ & $<0.57$ & $>1.25$ \\
& $211552.88+000115.5$ & $<3.3$ & $<3.2$ & $<3.3$ & \nodata & \nodata \\
& $232428.43+144324.3$ & $6.4^{+3.7}_{-2.5}$ & $4.1^{+3.2}_{-1.9}$ & $2.1^{+2.7}_{-1.3}$ & $0.52^{+0.81}_{-0.40} $ &  $1.38^{+1.31}_{-0.87} $ 
\enddata
\tablenotetext{a}{Errors on the \hbox{X-ray} counts were calculated using Poisson statistics corresponding to the 1$\sigma$ significance level according to Tables 1 and 2 of Gehrels (1986).}
\tablenotetext{b}{The band ratio is defined here as the number of hard-band counts divided by the number of soft-band counts. The errors on the band ratio correspond to the 1$\sigma$ significance level and were calculated using equation (1.31) in \S 1.7.3 of Lyons (1991). The band ratios for all of the {\it Chandra} objects observed in the same cycle can be directly compared with one another.}
\tablenotetext{c}{The effective power-law photon indices were calculated using the {\it Chandra} PIMMS tool (version 3.9$k$). The effect of the quantum efficiency decay over time at low energies of the ACIS detector was corrected for {\it Chandra} observed objects.}
\end{deluxetable}
\end{center}

% Table 5: X-ray flux and \alpha_ox
%\setcounter{table}{3}
%\clearpage
\begin{turnpage}
\begin{deluxetable}{ccccccccccccccc}
\tablecolumns{15} \tabletypesize{\scriptsize}
%\tablenum{5}
\tablewidth{0pt}
\tablecaption{X-ray, Optical, and Radio Properties\label{aox_table}}
\tablehead{
    \colhead{} 
    & \colhead{}                     
%    & \colhead{}    
    & \colhead{}        
    & \colhead{}        
    & \colhead{}            
    & \colhead{Count}
    & \colhead{}                       
    & \colhead{}                 
    & \colhead{log $L$}                 
    & \colhead{}
    & \colhead{log $L_{\nu}$}          
    & \colhead{} 
    & \colhead{}
    & \colhead{}
    & \colhead{} \\
    \colhead{} 
    & \colhead{Object Name (SDSS J)} 
%    & \colhead{$z$} 
    & \colhead{$m_{i}$\tablenotemark{a}} 
    & \colhead{$M_{i}$} 
    & \colhead{$N_{\rm H}$} 
    & \colhead{Rate\tablenotemark{b}} 
    & \colhead{$F_{0.5-2\;{\rm keV}}$\tablenotemark{c}} 
    & \colhead{$f_{\rm 2\;keV}$\tablenotemark{d}} 
    & \colhead{($2-10\;{\rm keV}$)} 
    & \colhead{$f_{2500\mbox{\rm~\scriptsize\AA}}$\tablenotemark{e}} 
    & \colhead{(2500 \AA)}  
    & \colhead{$\alpha_{\rm ox}$} 
    & \colhead{$\Delta \alpha_{\rm ox}$ $(\sigma)$\tablenotemark{f}}
    & \colhead{$f_{\rm x-weak}$\tablenotemark{g}}
    & \colhead{$\alpha_{\rm ro}$} \\
    \colhead{} 
    & \colhead{(1)} 
    & \colhead{(2)} 
    & \colhead{(3)} 
    & \colhead{(4)} 
    & \colhead{(5)} 
    & \colhead{(6)} 
    & \colhead{(7)} 
    & \colhead{(8)} 
    & \colhead{(9)} 
    & \colhead{(10)} 
    & \colhead{(11)} 
    & \colhead{(12)} 
    & \colhead{(13)} 
    & \colhead{(14)}
%    & \colhead{(15)}
}
\startdata
\multicolumn{2}{l}{{\it Chandra} Cycle 12 Objects} & & & \\
& $081250.79+522530.8$ & $17.99$ & $-26.18$ & $4.33$ & $0.76^{+0.72}_{-0.41}$ & $0.34$ & $1.08$ & $43.48$ & $2.16$ & $30.88$ & $-2.03$ & $-0.42$ $(2.10)$ & $12.43$ & $>-0.093$ \\
& $094533.98+100950.1$ & $17.44$ & $-27.54$ & $2.90$ & $<1.04$ & $<0.44$ & $<1.77$ & $<44.00$ & $3.42$ & $31.39$ & $<-2.03$ & $<-0.34$ $(2.32)$ & $>7.69$ & $>-0.074$ \\
& $110938.50+373611.7$ & $18.01$ & $-23.50$ & $1.57$ & $132.38^{+6.84}_{-6.51}$ & $54.61$ & $113.79$ & $44.55$ & $0.83$ & $29.42$ & $-1.10$ & $0.31$ $(1.90)$ & $0.16$ & $-0.180$ \\
& $125219.47+264053.9$ & $17.71$ & $-26.65$ & $0.75$ & $0.92^{+0.87}_{-0.49}$ & $0.37$ & $1.27$ & $43.64$ & $2.53$ & $31.04$ & $-2.03$ & $-0.39$ $(2.69)$ & $10.38$ & $-0.109$ \\
& $153044.08+231013.4$ & $17.53$ & $-27.13$ & $4.23$ & $35.07^{+3.75}_{-3.40}$ & $15.54$ & $55.76$ & $45.35$ & $3.32$ & $31.23$ & $-1.45$ & $0.22$ $(1.50)$ & $0.27$ & $-0.184$ \\
& $161245.68+511816.9$ & $17.56$ & $-27.33$ & $1.66$ & $9.48^{+2.05}_{-1.71}$ & $3.92$ & $15.17$ & $44.89$ & $3.34$ & $31.34$ & $-1.67$ & $0.02$ $(0.11)$ & $0.89$ & $>-0.054$ \\
\\
\multicolumn{2}{l}{Archival \xray\ Data Objects} & & & \\
& $101353.46+492758.1$ & $18.23$ & $-26.68$ & $0.79$ & $<0.85$ & $<0.44$ & $<1.72$ & $<43.97$ & $1.59$ & $31.04$ & $<-1.91$ & $<-0.26$ $(1.81)$ & $>4.76$ & $>-0.115$ \\
& $113900.55-020140.0$ & $18.88$ & $-26.43$ & $2.58$ & $0.27^{+0.21}_{-0.13}$ & $0.12$ & $0.52$ & $43.55$ & $0.83$ & $30.88$ & $-2.00$ & $-0.38$ $(1.91)$ & $9.77$ & $>-0.172$ \\
& $160410.22+432614.6$\tablenotemark{h} & $17.84$ & $-26.94$ & $1.22$ & $1.56^{+0.21}_{-0.19}$ & $0.99$ & $3.74$ & $44.29$ & $2.21$ & $31.13$ & $-1.83$ & $-0.18$ $(1.22)$ & $2.94$ & $>-0.091$ \\
& $211552.88+000115.5$ & $17.84$ & $-26.94$ & $6.17$ & $<0.33$ & $<0.14$ & $<0.73$ & $<43.97$ & $2.39$ & $31.55$ & $<-2.12$ & $<-0.41$ $(2.78)$ & $>11.70$ & $-0.138$ \\
& $232428.43+144324.3$ & $19.22$ & $-26.81$ & $4.26$ & $0.82^{+0.64}_{-0.39}$ & $0.33$ & $1.18$ & $43.73$ & $1.21$ & $30.80$ & $-1.92$ & $-0.32$ $(1.60)$ & $6.82$ & $>-0.145$ 
\enddata
\tablenotetext{a}{The apparent $i$-band magnitude using the BEST photometry of the SDSS DR7 quasar catalog.}
\tablenotetext{b}{The count rate in the observed-frame soft \hbox{X-ray} band ($0.5-2.0$ keV) in units of $10^{-3}$ ${\rm s}^{-1}$.}
\tablenotetext{c}{The Galactic absorption-corrected observed-frame flux between $0.5-2.0$ keV in units of $10^{-14}$ ergs cm$^{-2}$ s$^{-1}$.}
\tablenotetext{d}{The flux density at rest-frame 2 keV, in units of $10^{-32}$ ergs cm$^{-2}$ s$^{-1}$ Hz$^{-1}$.}
\tablenotetext{e}{The flux density at rest-frame 2500~\AA\ in units of 10$^{-27}$ ergs cm$^{-2}$ s$^{-1}$ Hz$^{-1}$.}
\tablenotetext{f}{$\Delta\alpha_{\rm ox}$: the difference between the measured $\alpha_{\rm ox}$ and the expected $\alpha_{\rm ox}$, defined by the $\alpha_{\rm ox}-L_{2500~{\rm \AA}}$ relation in equation (3) of Just et al.~(2007); the statistical significance of this difference, $\sigma$, is measured in units of the RMS $\alpha_{\rm ox}$ defined in Table 5 of Steffen et al.~(2006).}
\tablenotetext{g}{The factor of X-ray weakness compared to a typical radio-quiet quasar with similar optical/UV luminosity; see \S3.}
\tablenotetext{h}{The X-ray properties for J1604+4326 reported here are for the average of its two \chandra\ observations.}
\end{deluxetable}
\end{turnpage}

% Table 6: Two sample test results
\begin{deluxetable}{lclccc}
\tabletypesize{\footnotesize}
%\tablenum{6}
\tablecaption{Results of Peto-Prentice Tests\label{twost_table}}
\tablewidth{0pt}
\tablehead{
%\colhead{} & & \multicolumn{2}{c}{J0903+0708 excluded} & & \multicolumn{2}{c}{J0903+0708 included} \\
%\cline{3-4}\cline{6-7} \\
\colhead{Sample I (No. of sources)} & \colhead{vs.} & \colhead{Sample II (No. of sources)} & \colhead{} & \colhead{Statistic} & \colhead{Null-hypothesis Probability} 
%& & \colhead{Statistic} & \colhead{Null-hypothesis Probability}% & \colhead{ Mode($\lambda$1900\,\AA)} & \colhead{ Mode(Fe\,{\sc iii})}
}
\startdata
Low-$z$ RQ WLQs (11) & & RQ Sample B quasars (132) & & $4.982$ & $6.29\times10^{-7}$ \\%& & $8.779$ & $1.65\times10^{-18}$ \\
High-$z$ RQ WLQs (9) & & RQ Sample B quasars (132) & & $2.693$ & $7.08\times10^{-3}$ \\
%high-z WLQs vs. Sample B quasars & & $0.827$ & $0.408$ \\
Low-$z$ + High-$z$ RQ WLQs (20) & & RQ Sample B quasars (132) & & $4.589$ & $4.45\times10^{-6}$ \\%& & $4.702$ & $2.58\times10^{-6}$ \\
Low-$z$ RQ WLQs (11) & & High-$z$ RQ WLQs (9) & & $0.932$  & $0.351$ %& & $4.008$ & $6.12\times10^{-5}$ 
%logrank & & $5.415$ & $6.13\times10^{-8}$ & & $5.334$ & $9.61\times10^{-8}$ \\
%Peto-Peto & & $4.687$ & $2.77\times10^{-6}$ & & $4.724$ & $2.31\times10^{-6}$\\
\enddata
\tablecomments{See Feigelson \& Nelson (1985) for the detailed definition of the test statistic. The null-hypothesis probability was calculated from each test statistic using a Gaussian distribution, e.g., $1-P_G=6.29\times10^{-7}$, where $P_G$ is the cumulative Gaussian probability at $4.982\sigma$.}
\end{deluxetable}

%\clearpage
% Table 7: X-ray spectral analysis
\begin{deluxetable}{cccccccc}
\tablecolumns{8} \tabletypesize{\footnotesize}
%\tablenum{7}
\tablewidth{0pt}
\tablecaption{X-ray Spectral Analysis}
\tablehead{
\colhead{}
%    & \colhead{}
    & \colhead{}
    & \multicolumn{2}{c}{Power Law}
    & \colhead{}
    & \multicolumn{3}{c}{Power Law}\\
    \colhead{}
%    & \colhead{Total Full-Band}
    & \colhead{}
    & \multicolumn{2}{c}{with Galactic Absorption}
    & \colhead{}
    & \multicolumn{3}{c}{with Galactic and Intrinsic Absorption}
%    & \colhead{}
%    & \colhead{} 
    \\
    \cline{3-4}\cline{6-8}\\    
\colhead{Object Name} 
%    & \colhead{Counts} 
    & \colhead{}
    & \colhead{$\Gamma$} 
    & \colhead{$\chi^2/\nu$} 
    & \colhead{}
    & \colhead{$\Gamma$} 
    & \colhead{$N_H (10^{22} \ {\rm cm}^{-2})$}
    & \colhead{$\chi^2/\nu$}
}
\startdata
J1109+3736 & & $1.77^{+0.14}_{-0.14}$ & $42.09/23$ & & $1.79^{+0.26}_{-0.15}$ & $<0.14$ & $42.08/22$ \\
J1530+2310 & & $2.11^{+0.37}_{-0.34}$ & $14.18/9$ & & $2.18^{+0.97}_{-0.41}$ & $<2.67$ & $14.13/8$ \\
J1604+4326 / J1612+5118 & & $2.07^{+0.31}_{-0.30}$ & $77.38/105$\tablenotemark{a} & & $2.10^{+0.53}_{-0.33}$ & $<1.58$ & $77.38/105$\tablenotemark{a} 
\enddata
\tablenotetext{a}{The numbers here are $C/n$ instead of $\chi^2/\nu$, where $C$ is the $C$-statistic, and 
$n$ is the total number of spectral bins.}
\label{xspec_table}
\end{deluxetable}
%\clearpage
%\end{landscape}

% -----------------------------------------------------------------------------
% Figures
% -----------------------------------------------------------------------------

\clearpage

% Figure A1_4: SEDs
%\begin{figure*}[t]
%    \figurenum{A1}
%    \centering
%    \includegraphics[width=6.2in]{../figures/p10arqseds4.ps}
%%    \includegraphics[width=5.0in]{fig05bcolor.ps}   
%    \caption{\footnotesize{\it Continued.}
%             \label{p10ased4_fig}}
%\end{figure*}

% Figure A1_5: SEDs
%\begin{figure*}[t]
%    \figurenum{A1}
%    \centering
%    \includegraphics[width=6.2in]{../figures/p10arqseds5.ps}
%%    \includegraphics[width=5.0in]{fig05bcolor.ps}   
%    \caption{\footnotesize{\it Continued.}
%             \label{p10ased5_fig}}
%\end{figure*}

% Figure A1_6: SEDs
%\begin{figure*}[t]
%    \figurenum{A1}
%    \centering
%    \includegraphics[width=6.2in]{../figures/p10arqseds6.ps}
%%    \includegraphics[width=5.0in]{fig05bcolor.ps}   
%    \caption{\footnotesize{\it Continued.}
%             \label{p10ased6_fig}}
%\end{figure*}

% Figure 7 cont: SEDs
%\begin{figure*}[t]
%    \figurenum{\ref{sed1_fig}}
%    \centering
%    \includegraphics[width=6.3in]{../figures/seds2.ps}
%    \includegraphics[width=5.0in]{fig05bcolor.ps}   
%    \caption{\footnotesize{\it Continued.}}
%             \label{sed2_fig}}
%\end{figure*}

% Figure 4: alpha_ro vs. delta_alpha_ox
%\begin{figure*}[t]
%    \centering
%    \includegraphics[width=6.3in]{../figures/arodaox.ps}
%%    \includegraphics[width=5.0in]{fig05bcolor.ps}   
%    \caption{\footnotesize{$\alpha_{\rm ro}$--$\Delta\alpha_{\rm ox}$ diagram for \hbox{low-redshift} WLQs (colored open squares) and BL~Lac objects (black open circles). 
%             Symbols are similar to those in Fig.~\ref{aroaox_fig}.}
%             \label{arodaox_fig}}
%\end{figure*}


\begin{thebibliography}

\bibitem[Abazajian et al.(2009)]{2009ApJS..182..543A} Abazajian, K.~N., 
Adelman-McCarthy, J.~K., Ag{\"u}eros, M.~A., et al.\ 2009, \apjs, 182, 543 

\bibitem[Anderson et al.(2001)]{2001AJ....122..503A} Anderson, S.~F., Fan, 
X., Richards, G.~T., et al.\ 2001, \aj, 122, 503 

\bibitem[Arnaud (1996)]{1996ASPC..101...17A} Arnaud, K.~A., 1996, in ASP Conf. Ser. 101,
Astronomical Data Analysis Software and Systems V, ed. G.~H.~Jacoby \& J.~Barnes (San Francisco:ASP), 17

\bibitem[Avni(1976)]{1976ApJ...210..642A} Avni, Y.\ 1976, \apj, 210, 642

\bibitem[Baldwin(1977)]{1977ApJ...214..679B} Baldwin, J.~A.\ 1977, \apj,
214, 679

\bibitem[Barvainis et al.(2005)]{2005ApJ...618..108B} Barvainis, R., 
Leh{\'a}r, J., Birkinshaw, M., Falcke, H., 
\& Blundell, K.~M.\ 2005, \apj, 618, 108 

\bibitem[Becker et al.(1995)]{1995ApJ...450..559B} Becker, R.~H., White, R.~L.,
  \& Helfand, D.~J., 1995, \apj, 450, 559

\bibitem[Cash(1979)]{1979ApJ...228..939C} Cash, W.\ 1979, \apj, 228, 939

\bibitem[Collinge et al.(2005)]{2005AJ....129.2542C} Collinge, M.~J., 
Strauss, M.~A., Hall, P.~B., et al.\ 2005, \aj, 129, 2542 

\bibitem[Diamond-Stanic et al.(2009)]{2009ApJ...699..782D} Diamond-Stanic, 
A.~M., Fan, X., Brandt, W.~N., et al.\ 2009, \apj, 699, 782

\bibitem[Donato et al.(2005)]{2005A&A...433.1163D} Donato, D., Sambruna, R.~M., 
\& Gliozzi, M.\ 2005, \aap, 433, 1163

\bibitem[Efron(1979)]{1979...7...1} Efron, B., 1979, The Annals of Statistics, 7, 1

%\bibitem[Elvis(2000)]{2000ApJ...545...63E} Elvis, M.\ 2000, \apj, 545, 63 

\bibitem[Eracleous et al.(2009)]{2009NewAR..53..133E} Eracleous, M., Lewis,
K.~T., \& Flohic, H.~M.~L.~G.\ 2009, New A Rev., 53, 133

\bibitem[Falcke et al.(1996)]{1996ApJ...471..106F} Falcke, H., Sherwood, 
W., \& Patnaik, A.~R.\ 1996, \apj, 471, 106 

\bibitem[Fan et al.(1999)]{1999ApJ...526L..57F} Fan, X., Strauss, M.~A., 
Gunn, J.~E., et al.\ 1999, \apjl, 526, L57 

\bibitem[Fan et al.(2006)]{2006AJ....131.1203F} Fan, X., Strauss, M.~A., 
Richards, G.~T., et al.\ 2006, \aj, 131, 1203 

\bibitem[Feigelson \& Nelson(1985)]{1985ApJ...293..192F} Feigelson, E.~D., 
\& Nelson, P.~I.\ 1985, \apj, 293, 192

\bibitem[Fisher(1922)]{1922......} Fisher, R.~A. 1922, Journal of the Royal Statistical Society, 85(1), 87

\bibitem[Freeman et al.(2002)]{2002ApJS..138..185F} Freeman, P.~E.,
Kashyap, V., Rosner, R., \& Lamb, D.~Q.\ 2002, \apjs, 138, 185

\bibitem[Garmire et al.(2003)]{2003SPIE.4851...28G} Garmire, G.~P., Bautz,
M.~W., Ford, P.~G., Nousek, J.~A.,
\& Ricker, G.~R., Jr.\ 2003, \procspie, 4851, 28

\bibitem[Gehrels(1986)]{1986ApJ...303..336G} Gehrels, N. 1986,
  \apj, 303, 336

\bibitem[Gibson et al.(2008)]{2008ApJ...685..773G} Gibson, R.~R., Brandt,
W.~N., \& Schneider, D.~P.\ 2008, \apj, 685, 773

\bibitem[Gibson et al.(2009)]{2009ApJ...692..758G} Gibson, R.~R., Jiang, 
L., Brandt, W.~N., et al.\ 2009, \apj, 692, 758 

%\bibitem[Hawkins(2004)]{2004A&A...424..519H} Hawkins, M.~R.~S.\ 2004, \aap, 424, 519

\bibitem[Heidt 
\& Nilsson(2011)]{2011A&A...529A.162H} Heidt, J., \& Nilsson, K.\ 2011, \aap, 529, A162

\bibitem[Hewett 
\& Wild(2010)]{2010MNRAS.405.2302H} Hewett, P.~C., \& Wild, V.\ 2010, \mnras, 405, 2302

\bibitem[Hill et al.(1998)]{1998SPIE.3355..375H} Hill, G.~J., Nicklas, 
H.~E., MacQueen, P.~J., et al.\ 1998, \procspie, 3355, 375 

\bibitem[Hryniewicz et al.(2010)]{2010MNRAS.404.2028H} Hryniewicz, K.,
Czerny, B., Niko{\l}ajuk, M., \& Kuraszkiewicz, J.\ 2010, \mnras, 404, 2028

\bibitem[Isobe et al.(1990)]{1990ApJ...364..104I} Isobe, T., Feigelson,
E.~D., Akritas, M.~G., \& Babu, G.~J.\ 1990, \apj, 364, 104

\bibitem[Jester et al.(2005)]{2005AJ....130..873J} Jester, S., Schneider, 
D.~P., Richards, G.~T., et al.\ 2005, \aj, 130, 873 

\bibitem[Just et al.(2007)]{2007ApJ...665.1004J} Just, D.~W., Brandt, 
W.~N., Shemmer, O., et al.\ 2007, \apj, 665, 1004 

\bibitem[Kellermann et al.(1989)]{1989AJ.....98.1195K} Kellermann, K.~I.,
Sramek, R., Schmidt, M., Shaffer, D.~B., \& Green, R.\ 1989, \aj,
98, 1195

\bibitem[Kelly(2007)]{2007ApJ...665.1489K} Kelly, B.~C.\ 2007, \apj, 665,
1489

\bibitem[Komatsu et al.(2009)]{2009ApJS..180..330K} Komatsu, E., Dunkley, 
J., Nolta, M.~R., et al.\ 2009, \apjs, 180, 330 

\bibitem[Kraft et al.(1991)]{1991ApJ...374..344K} Kraft, R.~P., Burrows,
D.~N., \& Nousek, J.~A.\ 1991, \apj, 374, 344

\bibitem[Lane et al.(2011)]{2011ApJ...743..163L} Lane, R.~A., Shemmer, O., 
Diamond-Stanic, A.~M., et al.\ 2011, \apj, 743, 163 

\bibitem[Laor 
\& Davis(2011)]{2011arXiv1106.4969L} Laor, A., \& Davis, S.~W.\ 2011, \mnras, 417, 681

\bibitem[Latta(1981)]{1981JASA...76...713L}Latta, R. B., 1981, J. Am. Stat. Assoc.,
76, 713

\bibitem[Lavalley et al.~(1992)]{1992ASPC...25..245L} Lavalley, M., Isobe, T.,
  Feigelson, E., 1992, in ASP Conf. Ser. 25, Astronomical Data Analysis Software
  and Systems I, ed. D.~M. Worrall, C. Biemesderfer, \& J. Barnes (San
  Francisco, CA: ASP), 245

\bibitem[Leighly(2004)]{2004ApJ...611..125L} Leighly, K.~M.\ 2004, \apj, 
611, 125

\bibitem[Leighly et al.(2007)]{2007ApJ...663..103L} Leighly, K.~M., 
Halpern, J.~P., Jenkins, E.~B., et al.\ 2007a, \apj, 663, 103 

\bibitem[Leighly et al.(2007)]{2007ApJS..173....1L} Leighly, K.~M.,
Halpern, J.~P., Jenkins, E.~B., \& Casebeer, D.\ 2007b, \apjs, 173, 1

\bibitem[Liu 
\& Zhang(2011)]{2011ApJ...728L..44L} Liu, Y., \& Zhang, S.~N.\ 2011, \apjl, 728, L44

\bibitem[Lyons(1991)]{1991...............} Lyons, L. 1991, Data Analysis
  for Physical Science Students (Cambridge: Cambridge Univ. Press)

\bibitem[Londish et al.(2004)]{2004MNRAS.352..903L} Londish, D., Heidt, J.,
Boyle, B.~J., Croom, S.~M.,
\& Kedziora-Chudczer, L.\ 2004, \mnras, 352, 903

\bibitem[Martin et al.(2005)]{2005ApJ...619L...1M} Martin, D.~C., Fanson, 
J., Schiminovich, D., et al.\ 2005, \apjl, 619, L1 

\bibitem[McDowell et al.(1995)]{1995ApJ...450..585M} McDowell, J.~C., 
Canizares, C., Elvis, M., et al.\ 1995, \apj, 450, 585 

\bibitem[McMahon et al.(2002)]{2002ApJS..143....1M} McMahon, R.~G., White,
R.~L., Helfand, D.~J., \& Becker, R.~H.\ 2002, \apjs, 143, 1

\bibitem[Meusinger et al.(2011)]{2011A&A...525A..37M} Meusinger, H., 
Hinze, A., \& de Hoon, A.\ 2011, \aap, 525, A37

\bibitem[Morrison \& McCammon(1983)]{1983ApJ...270..119M} Morrison, R., 
\& McCammon, D.\ 1983, \apj, 270, 119

\bibitem[Murray 
\& Chiang(1997)]{1997ApJ...474...91M} Murray, N., \& Chiang, J.\ 1997, \apj, 474, 91 

\bibitem[Nestor et al.(2008)]{2008MNRAS.386.2055N} Nestor, D., Hamann, F., 
\& Rodriguez Hidalgo, P.\ 2008, \mnras, 386, 2055 

\bibitem[Nieppola et al.(2006)]{2006A&A...445..441N} Nieppola, E., 
Tornikoski, M., \& Valtaoja, E.\ 2006, \aap, 445, 441

\bibitem[Nousek \& Shue(1989)]{1989ApJ...342.1207N} Nousek, J.~A., \& Shue, D.~R.\ 1989, \apj, 342, 1207

\bibitem[Plotkin et al.(2010)]{2010AJ....139..390P} Plotkin, R.~M., 
Anderson, S.~F., Brandt, W.~N., et al.\ 2010a, \aj, 139, 390 

\bibitem[Plotkin et al.(2010)]{2010ApJ...721..562P} Plotkin, R.~M., 
Anderson, S.~F., Brandt, W.~N., et al.\ 2010b, \apj, 721, 562 

\bibitem[Plotkin et al.(2011)]{2011......} Plotkin, R.~M., Anderson, S.~F., Brandt, W.~N., Markoff, S., 
Shemmer, O., \& Wu, J.\ 2011, \apj\ Letters, submitted

\bibitem[Prochaska et al.(2008)]{2008ApJ...675.1002P} Prochaska, J.~X., 
Hennawi, J.~F., \& Herbert-Fort, S.\ 2008, \apj, 675, 1002 

\bibitem[Proga et al.(2000)]{2000ApJ...543..686P} Proga, D., Stone, J.~M., 
\& Kallman, T.~R.\ 2000, \apj, 543, 686 

\bibitem[Ramsey et al.(1998)]{1998SPIE.3352...34R} Ramsey, L.~W., Adams, 
M.~T., Barnes, T.~G., et al.\ 1998, \procspie, 3352, 34 

\bibitem[Richards et al.(2002)]{2002AJ....123.2945R} Richards, G.~T., Fan, 
X., Newberg, H.~J., et al.\ 2002, \aj, 123, 2945 

\bibitem[Richards et al.(2006)]{2006ApJS..166..470R} Richards, G.~T., Lacy, 
M., Storrie-Lombardi, L.~J., et al.\ 2006, \apjs, 166, 470 

\bibitem[Richards et al.(2011)]{2011AJ....141..167R} Richards, G.~T., 
Kruczek, N.~E., Gallagher, S.~C., et al.\ 2011, \aj, 141, 167 

\bibitem[Schneider et al.(2010)]{2010AJ....139.2360S} Schneider, D.~P., 
Richards, G.~T., Hall, P.~B., et al.\ 2010, \aj, 139, 2360 

\bibitem[Sesar et al.(2006)]{2006AJ....131.2801S} Sesar, B., Svilkovi{\'c}, 
D., Ivezi{\'c}, {\v Z}., et al.\ 2006, \aj, 131, 2801 

\bibitem[Shemmer et al.(2006)]{2006ApJ...644...86S} Shemmer, O., Brandt, 
W.~N., Schneider, D.~P., et al.\ 2006, \apj, 644, 86 

\bibitem[Shemmer et al.(2008)]{2008ApJ...682...81S} Shemmer, O., Brandt,
W.~N., Netzer, H., Maiolino, R., \& Kaspi, S.\ 2008, \apj, 682, 81

\bibitem[Shemmer et al.(2009)]{2009ApJ...696..580S} Shemmer, O., Brandt, 
W.~N., Anderson, S.~F., et al.\ 2009, \apj, 696, 580 

\bibitem[Shemmer et al.(2010)]{2010ApJ...722L.152S} Shemmer, O., 
Trakhtenbrot, B., Anderson, S.~F., et al.\ 2010, \apjl, 722, L152 

\bibitem[Shen et al.(2008)]{2008ApJ...680..169S} Shen, Y., Greene, J.~E., 
Strauss, M.~A., Richards, G.~T., \& Schneider, D.~P.\ 2008, \apj, 680, 169

\bibitem[Shen et al.(2011)]{2011ApJS..194...45S} Shen, Y., Richards, G.~T., 
Strauss, M.~A., et al.\ 2011, \apjs, 194, 45 

\bibitem[Skrutskie et al.(2006)]{2006AJ....131.1163S} Skrutskie, M.~F., 
Cutri, R.~M., Stiening, R., et al.\ 2006, \aj, 131, 1163 

\bibitem[Smith et al.(2007)]{2007ApJ...663..118S} Smith, P.~S., Williams,
G.~G., Schmidt, G.~D., Diamond-Stanic, A.~M.,
\& Means, D.~L.\ 2007, \apj, 663, 118

\bibitem[Steffen et al.(2006)]{2006AJ....131.2826S} Steffen, A.~T., 
Strateva, I., Brandt, W.~N., et al.\ 2006, \aj, 131, 2826 

%\bibitem[Trump et al.(2009)]{2009ApJ...706..797T} Trump, J.~R., et al.\
%2009, \apj, 706, 797
%\bibitem[Trump et al.(2011)]{2011ApJ...732...23T} Trump, J.~R., et al.\
%2011, \apj, 732, 23

\bibitem[Vanden Berk et al.(2001)]{2001AJ....122..549V} Vanden Berk, D.~E., 
Richards, G.~T., Bauer, A., et al.\ 2001, \aj, 122, 549 

%\bibitem[Werner et al.(2004)]{2004ApJS..154....1W} Werner, M.~W., et al.\
%2004, \apjs, 154, 1

%\bibitem[White et al.(2010)]{2010arXiv1001.2843W} White, N.~E., Parmar, A.,
%Kunieda, H., Nandra, K., Ohashi, T.,
%\& Bookbinder, J.\ 2010, in AIP Conf. Proc. 1248, \xray\ Astronomy 2009: Present
%Status, Multi-wavelength Approach and Future Perspectives, ed. A.~Comastri, L.~Angelini,
%\& M.~Cappi (Melville, NY:AIP), 561

\bibitem[Wright et al.(2010)]{2010AJ....140.1868W} Wright, E.~L., 
Eisenhardt, P.~R.~M., Mainzer, A.~K., et al.\ 2010, \aj, 140, 1868 

\bibitem[Wu et al.(2011)]{2011ApJ...736...28W} Wu, J., Brandt, W.~N., Hall, 
P.~B., et al.\ 2011, \apj, 736, 28 

\bibitem[Voges et 
al.(1999)]{1999A&A...349..389V} Voges, W., Aschenbach, B., Boller, T., et al.\ 1999, \aap, 349, 389 

\bibitem[York et al.(2000)]{2000AJ....120.1579Y} York, D.~G., Adelman, J., 
Anderson, J.~E., Jr., et al.\ 2000, \aj, 120, 1579 

\end{thebibliography}
\end{document}